\documentclass[acmtog]{acmart}

\usepackage{booktabs} 
\usepackage{todonotes, diagbox, multirow}
\definecolor{custompink}{RGB}{237,2,140}

\usepackage[linesnumbered,ruled,vlined]{algorithm2e} 

\SetAlFnt{\small}
\SetAlCapFnt{\small}
\SetAlCapNameFnt{\small}
\SetAlCapHSkip{0pt}

\definecolor{darkgreen}{rgb}{0.0, 0.4, 0.0} 
\newcommand*{\Method}{\textit{UniRig}~}
\newcommand{\Dataset}{\textit{Rig-XL}~}
\newcommand{\cal}{\mathcal}
\newcommand{\bb}{\mathbb}

\usepackage{pifont}
\newcommand{\ccheck}{\textcolor{darkgreen}{\ding{51}}}%
\newcommand{\ctimes}{\textcolor{red}{\text{\ding{55}}}}

\newcommand{\cyp}[1]{\textcolor{black}{#1}}
\newcommand{\zjp}[1]{\textcolor{black}{#1}}

\begin{document}

\title{One Model to Rig Them All: Diverse Skeleton Rigging with \emph{UniRig} }

\author{Jia-Peng Zhang}
\email{zjp24@mails.tsinghua.edu.cn}
\affiliation{%
 \institution{BNRist, Department of Computer Science and Technology, Tsinghua University}
 \city{Beijing}
 \country{China}}
\author{Cheng-Feng Pu}
\email{pcf22@mails.tsinghua.edu.cn}
\affiliation{%
 \institution{Zhili College, Tsinghua University}
 \city{Beijing}
 \country{China}}
\author{Meng-Hao Guo}
\email{gmh20@mails.tsinghua.edu.cn}
\affiliation{%
 \institution{BNRist, Department of Computer Science and Technology, Tsinghua University}
 \city{Beijing}
 \country{China}}
\author{Yan-Pei Cao}
\email{caoyanpei@gmail.com}
\affiliation{%
\institution{VAST}
\city{Beijing}
\country{China}}
\author{Shi-Min Hu}
\email{shimin@tsinghua.edu.cn}
\affiliation{%
\institution{BNRist, Department of Computer Science and Technology, Tsinghua University}
\city{Beijing}
\country{China}}


\begin{abstract}

The rapid evolution of 3D content creation, encompassing both AI-powered methods and traditional workflows, is driving an unprecedented demand for automated rigging solutions that can keep pace with the increasing complexity and diversity of 3D models. We introduce \emph{UniRig}, a novel, unified framework for automatic skeletal rigging that leverages the power of large autoregressive models and a bone-point cross-attention mechanism to generate both high-quality skeletons and skinning weights. Unlike previous methods that struggle with complex or non-standard topologies, \Method accurately predicts topologically valid skeleton structures thanks to a new \emph{Skeleton Tree Tokenization} method that efficiently encodes hierarchical relationships within the skeleton. To train and evaluate UniRig, we present \emph{Rig-XL}, a new large-scale dataset of over 14,000 rigged 3D models spanning a wide range of categories. \Method significantly outperforms state-of-the-art academic and commercial methods, achieving a 215\% improvement in rigging accuracy and a 194\% improvement in motion accuracy on challenging datasets. Our method works seamlessly across diverse object categories, from detailed anime characters to complex organic and inorganic structures, demonstrating its versatility and robustness. By automating the tedious and time-consuming rigging process, \Method has the potential to speed up animation pipelines with unprecedented ease and efficiency. \textcolor{custompink}{Project Page: \href{https://zjp-shadow.github.io/works/UniRig/}{https://zjp-shadow.github.io/works/UniRig/} }
\end{abstract}




\begin{teaserfigure}
  \includegraphics[width=\textwidth]{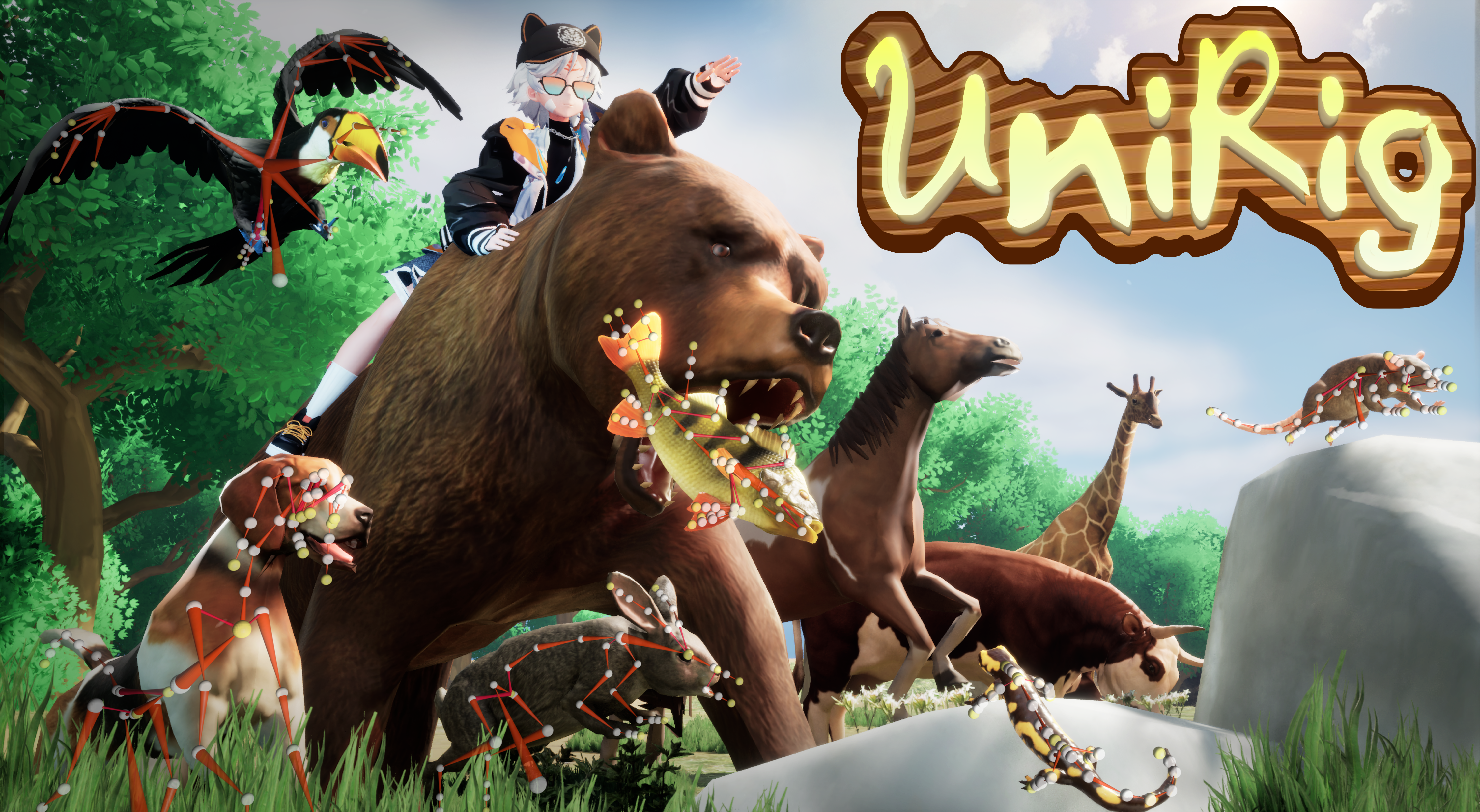}
  \vspace{-4mm}
  \caption{Diverse 3D models rigged using \emph{UniRig}. The models, spanning various categories including animals, humans, and fictional characters, demonstrate the versatility of our method. Selected models are visualized with their predicted skeletons. $\copyright$ Tira}
  \label{fig:teaser}
\end{teaserfigure}
\maketitle

\section{Introduction}

 \begin{table*}[!t]
 \caption{\textbf{Comparison of \Method with Prior Work in Automatic Rigging.} $\ast$ Tripo supports only human and quadruped categories. $\dagger$ Inference time depends on the number of bones and the complexity of the model.}
 \resizebox{\linewidth}{!}{
 \renewcommand{\tabcolsep}{5mm}
     \centering
     \begin{tabular}{l|c|c|c|c|c}
              \hline
         {Method} &  {Template Based} & {Template Free} & {Automation Level} & {Multi Categories}  & {Cost Time}\\
         \hline
         RigNet \cite{xu2020rignet} & \ctimes & \ccheck & Automated & \ccheck & $1 \text{s} \sim 20 \text{min} ^{\dagger}$\\
         NBS \cite{li2021learning} & \ccheck & \ctimes & Automated & \ctimes & $1$ s\\
         TaRig \cite{ma2023tarig} & \ccheck & \ccheck & Automated & \ctimes & $30$ s\\
         Anything World \cite{anythingworld} & \ccheck & \ccheck & Semi-Automated & \ccheck & $5$ min\\
         Tripo \cite{tripoai} & \ccheck  & \ccheck & Automated & \ccheck$^{\ast}$ & $2$ min\\
         Meshy \cite{meshy} &  \ccheck & \ctimes & Semi-Automated & \ctimes & $1 \sim 2$ min\\
         Accurig \cite{accurig} & \ccheck & \ctimes & Semi-Automated & \ctimes & $1$ min\\
         \Method(Ours) & \ccheck & \ccheck & Automated & \ccheck & $1\sim 5$ s\\
                  \hline
     \end{tabular}
     }
     \vspace{-4mm}
     \label{tab:Advatanges}
     
 \end{table*}

The rapid advancements in AI-driven 3D content creation \cite{poole2022dreamfusion,zhang2024clay,yu2024texgen,siddiqui2024meshgpt,peng2024charactergen,holden2017phase} are revolutionizing computer graphics, enabling the generation of complex 3D models at an unprecedented scale and speed. This surge in automatically generated 3D content has created a critical need for efficient and robust rigging solutions, as manual rigging remains a time-consuming and expertise-intensive bottleneck in the animation pipeline. While skeletal animation has long been a cornerstone of 3D animation, traditional rigging techniques often require expert knowledge and hours of time to complete for a single model.

The rise of deep learning has spurred the development of automatic rigging methods, offering the potential to dramatically accelerate this process. Existing methods can be broadly categorized as template-based or template-free. Template-based approaches \cite{chu2024humanrig,li2021learning,liu2019neuroskinning} rely on predefined skeleton templates (e.g., SMPL \cite{loper2023smpl}) and achieve high accuracy in predicting bone positions within those templates. However, they are limited to specific skeleton topologies and struggle with models that deviate from the predefined templates. Template-free methods, such as RigNet \cite{xu2020rignet}, offer greater flexibility by predicting skeleton joints and their connectivity without relying on a template. However, these methods often produce less stable results and may generate topologically implausible skeletons. Furthermore, retargeting motion to these generated skeletons can be challenging.

Another line of research has explored skeleton-free mesh deformation \cite{aigerman2022neural, wang2023zero, liao2022skeleton}, which bypasses the need for explicit skeleton structures. While these methods offer intriguing possibilities, they often rely heavily on existing motion data, making them less generalizable to new and unseen motions. They also tend to be less compatible with established industry pipelines that rely on skeletal animation. Fully neural network-based methods can be computationally expensive, limiting their applicability in resource-constrained scenarios.

Despite these advancements, existing automatic rigging techniques still fall short in addressing the growing demand for rigging diverse 3D models. As highlighted in Table \ref{tab:Advatanges}, many methods are limited to specific model categories, struggle with complex topologies, or rely on manual intervention. To overcome these limitations, we propose \emph{UniRig}, a novel learning-based framework for automatic rigging of diverse 3D models.

A key challenge in automatic rigging is the inherent complexity of representing and generating valid skeleton structures. They possess a hierarchical tree structure with complex interdependencies between joints. Previous template-free methods often struggled to accurately capture these topological constraints, leading to unstable or unrealistic skeletons. \Method addresses this challenge by leveraging the power of autoregressive models, which excel at capturing sequential dependencies and generating structured outputs.
Specifically, \Method employs an autoregressive model to predict the skeleton tree in a topologically sorted order, ensuring the generation of valid and well-structured skeletons. This is enabled by a novel \emph{Skeleton Tree Tokenization} method that efficiently encodes the skeleton's hierarchical structure into a sequence of tokens. This tokenization scheme is designed to explicitly represent the parent-child relationships within the skeleton tree, guiding the autoregressive model to produce topologically sound outputs. Furthermore, the tokenization incorporates information about specific bone types (e.g., spring bones, template bones), facilitating downstream tasks such as motion retargeting. \Method also leverages a Bone-Point Cross Attention mechanism to accurately predict skinning weights, capturing the complex relationships between the generated skeleton and the input mesh.

To train \Method, we curated \Dataset, a new large-scale dataset of over 14,000 3D models with diverse skeletal structures and corresponding skinning weights. \Dataset significantly expands upon existing datasets in terms of both size and diversity, enabling us to train a highly generalizable model. We also leverage \emph{VRoid}, a dataset of anime-style characters, to refine our model's ability to handle detailed character models.

Our contributions can be summarized as follows:

\begin{itemize}
    \item We propose a novel Skeleton Tree Tokenization method that efficiently encodes skeletal structures, enabling the autoregressive model to generate topologically valid and well-structured skeletons.
    \item We curate and present \Dataset, a new large-scale and diverse dataset of 3D rigged models. This dataset has been carefully cleaned and provides a high-quality, generalized resource for subsequent auto-rigging tasks.
    \item We introduce \emph{UniRig}, a unified framework for automatic rigging that combines an autoregressive model for skeleton prediction with a Bone-Point Cross Attention mechanism for skin weight prediction. We demonstrate that \Method achieves state-of-the-art results in both skeleton prediction and skinning weight prediction, outperforming existing methods on a wide range of object categories and skeletal structures.
\end{itemize}

\section{Related Works}

\subsection{Data-Driven Mesh Deformation Transfer}

The skeleton animation system \cite{marr1978representation} is a foundational technique in computer graphics animation. However, some studies \cite{xu2020rignet,zhang2023tapmo} suggest that mastering rigging methods can be challenging for non-experts. Recently, in the field of character animation, driven by advancements in deep learning and the availability of numerous datasets \cite{xu2019predicting,Models-Resource,blackman2014rigging,chu2024humanrig}, mesh-deformation methods that bypass traditional rigging processes have emerged. These methods can be broadly classified into two categories, as outlined below:

\subsubsection{Skeleton-free Mesh Deformation\label{sec:related_1}}
Some methods \cite{zhang2024skinned, wang2023hmc} bypass the explicit representation of a skeleton and instead learn to directly deform the mesh based on input parameters or learned motion patterns.

SfPT \cite{liao2022skeleton} introduces a center-based Linear Blend Skinning (LBS) \cite{kavan2007skinning} method and constructs a Pose Transfer Network that leverages deep learning to facilitate motion transfer across characters. Building on this approach, HMC \cite{wang2023hmc} proposes an iterative method for mesh deformation prediction, improving accuracy by refining predictions from coarse to fine levels. Tapmo \cite{zhang2023tapmo}, inspired by SfPT, employs a Mesh Handle Predictor and Motion Diffusion to generate motion sequences and retarget them to diverse characters. 

\subsubsection{Vertex Displacement Prediction\label{sec:related_2}}

Another approach is to drive entirely through neural networks, and some research\cite{groueix20183d, yu2025mesh2animation} efforts have also explored this.
\cite{wang2020neural} introduced the first neural pose transfer model for human characters. \cite{gao2018automatic} proposed a VAE-Cycle-GAN framework that uses cycle consistency loss between source and target characters to predict mesh deformation automatically. ZPT \cite{wang2023zero} develops a correspondence-aware shape understanding module to enable zero-shot retargeting of stylized characters.

\vspace{1mm}
\zjp{While promising, the skeleton-free and direct vertex displacement approaches described in Sections \ref{sec:related_1} and \ref{sec:related_2} face challenges in integrating with established industry workflows, which heavily rely on traditional skeletal rigging and animation systems.}

\vspace{-1mm}
\subsection{Automatic Rigging Methods}

Automatic rigging aims to automate the process of creating a skeleton and associating it with a 3D mesh. Existing approaches can be categorized as either traditional geometry-based methods or more recent deep learning-based techniques.

\subsubsection{Traditional Geometric Methods}

Early methods \cite{amenta1998surface,tagliasacchi2009curve} relied on traditional geometric features to predict skeletons without requiring data. Pinocchio \cite{baran2007automatic} approximates the medial surface using signed distance fields and optimizes skeleton embedding via discrete penalty functions. Geometric techniques like Voxel Cores \cite{yan2018voxel} and Erosion Thickness \cite{yan2016erosion}, which fit medial axes and surfaces, also use these structures to drive 3D meshes in a manner similar to skeletons. Although these traditional methods can effectively handle objects with complex topologies, they often require significant manual intervention within industrial pipelines. For instance, tools such as LazyBones \cite{lazy-bones}, based on medial axis fitting, still necessitate considerable animator input to fine-tune skeletons before they can be used in production.

\subsubsection{Deep Learning Algorithms}

With the rapid advancement of deep learning, several data-driven auto-rigging methods \cite{ma2023tarig,liu2019neuroskinning,wang2025towards} have emerged in animation. RigNet \cite{xu2020rignet} is a notable example, which uses animated character data to predict joint heatmaps and employs the Minimum Spanning Tree algorithm to connect joints, achieving automatic skeletal rigging for various objects. MoRig \cite{xu2022morig} enhances RigNet by using a motion encoder to capture geometric features, improving both accuracy and precision in the joint extraction process. To address the artifacts commonly seen in LBS-based systems, Neural Blend Shapes \cite{li2021learning} introduces a residual deformation branch to improve deformation quality at joint regions. DRiVE \cite{sun2024drive} applies Gaussian Splatting conditioned Diffusion to predict joint positions. \zjp{However, these methods often require a separate step to infer bone connectivity from the predicted joints, which can introduce topological errors.}

Many existing deep learning-based methods suffer from limitations that hinder their widespread applicability. Some methods are restricted to specific skeleton topologies (e.g., humanoids), while others rely on indirect prediction of bone connections, leading to potential topological errors. These methods often struggle to balance flexibility with stability and precision. Our work addresses these limitations by leveraging an autoregressive model for skeleton prediction. This approach is inspired by recent advancements in 3D autoregressive generation \cite{hao2024meshtron, siddiqui2024meshgpt, chen2024meshanything} that have shown promise in modeling 3D shapes using tokenization and sequential prediction.

\begin{figure*}[!t]
    \centering
    \includegraphics[width=0.95\linewidth]{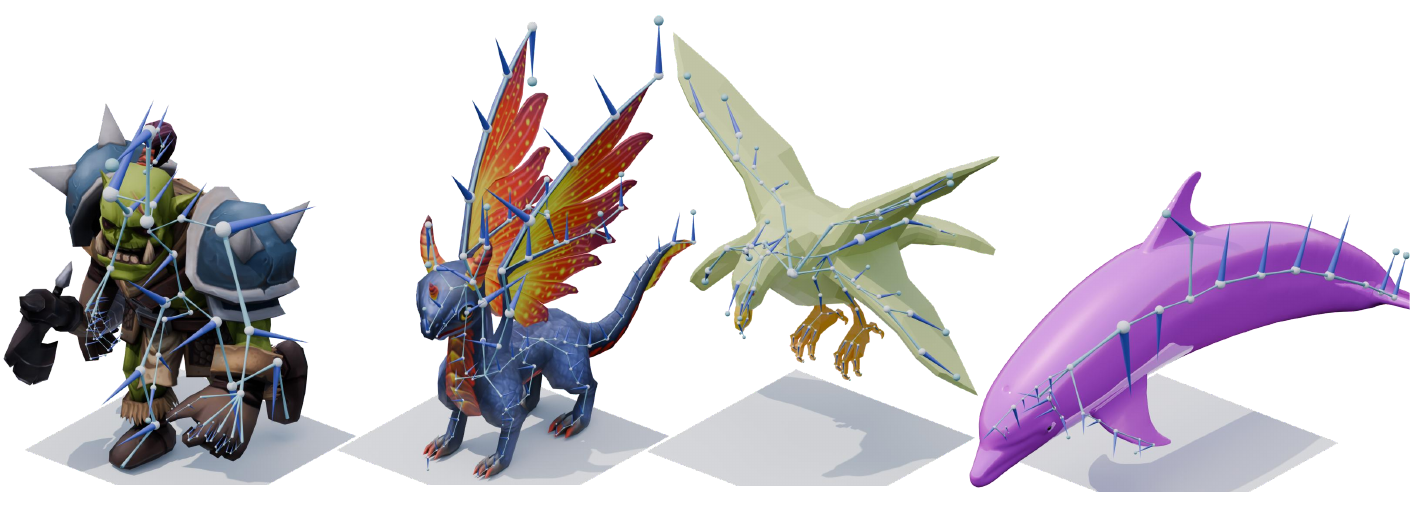}
    \vspace{-5mm}
    \caption{Examples from \Dataset, demonstrating well-defined skeleton structures.}
    \label{fig:enter-label}
\end{figure*}

\vspace{-1mm}
\section{Overview}

\cyp{The core challenge in automated skeletal rigging lies in accurately predicting both a plausible skeleton structure and the associated skinning weights that define mesh deformation. Previous methods often struggle with the diversity of 3D model topologies, requiring manual intervention or specialized approaches for different categories. To address this, we propose \Method, a unified learning-based framework for rigging diverse 3D models. \Method employs a novel paradigm that effectively combines two learned models into a single streamlined rigging process. It consists of two key stages: (1) autoregressive skeleton tree prediction from an input mesh (Section~\ref{sec:ar-skeleton}),  leveraging a novel tokenization method for efficient processing, and (2) efficient per-point skin weight prediction conditioned on the predicted skeleton, using a Bone-Point Cross Attention mechanism (Section~\ref{sec:skin_pred}).}

\cyp{To train and evaluate \Method, we introduce two datasets: VRoid (Section~\ref{sec:dataset-vroid}), a collection of anime-style 3D human models, and \Dataset (Section~\ref{sec:dataset-rigxl}), a new large-scale dataset spanning over 14,000 diverse and high-quality 3D models. VRoid helps refine our method's ability to model fine details, while \Dataset ensures generalizability across a wide range of object categories.}

\cyp{We evaluate UniRig's performance through extensive experiments (Section~\ref{sec:experiments}), comparing it against state-of-the-art methods and commercial tools. Our results demonstrate significant improvements in both rigging accuracy and animation fidelity. We further showcase UniRig's practical applications in human-assisted auto-rigging and character animation (Section~\ref{sec:Application}). Finally, we discuss limitations and future work (Section~\ref{sec:limitation}).}

\section{Dataset\label{sec:dataset}}

\subsection{VRoid Dataset Curation\label{sec:dataset-vroid}}

\cyp{To facilitate the development of detailed and expressive skeletal rigs, particularly for human-like characters, we have curated a dataset of $2,061$ anime-style 3D models from VRoidHub~\cite{isozaki2021vroid}.}

\cyp{This dataset, which we refer to as \emph{VRoid}, is valuable for training models capable of capturing the nuances of character animation, including subtle movements and deformations. It complements our larger and more diverse Rig-XL dataset (Section~\ref{sec:dataset-rigxl}) by providing a focused collection of models with detailed skeletal structures.}

\cyp{The VRoid dataset was compiled by first filtering the available models on VRoidHub based on the number of bones. These models were further refined through a manual selection process to ensure data quality and consistency in skeletal structure and to eliminate models with incomplete or improperly defined rigs.}

\subsubsection{VRM Format}

\cyp{The models in the VRoid dataset are provided in the VRM format, a standardized file format for 3D avatars used in virtual reality applications. A key feature of the VRM format is its standardized humanoid skeleton definition, which is compatible with the widely used Mixamo~\cite{blackman2014rigging} skeleton. This standardization simplifies the process of retargeting and animating these models. Furthermore, the VRM format supports \emph{spring bones}~\cite{isozaki2021vroid}, which are special bones that simulate physical interactions like swaying and bouncing. These spring bones are crucial for creating realistic and dynamic motion in parts of the model such as hair, clothing, and tails, as demonstrated in Figure~\ref{fig:compare_spring}. The behavior of these spring bones is governed by a physics simulation, detailed in Section~~\ref{sec:physics}. The inclusion of spring bones in the VRoid dataset allows our model to learn to generate rigs that support these dynamic effects, leading to more lifelike and engaging animations.}

\subsection{\Dataset Dataset Curation\label{sec:dataset-rigxl}}

\cyp{To train a truly generalizable rigging model capable of handling diverse object categories, a large-scale dataset with varied skeletal structures and complete skinning weights is essential. To this end, we curated \Dataset, a new dataset derived from the Objaverse-XL dataset~\cite{deitke2024objaverse}, which contains over 10 million 3D models. While Objaverse-XL is a valuable resource, it primarily consists of static objects and lacks the consistent skeletal structure and skinning weight information required for our task. We address this by filtering and refining the dataset.}

\cyp{We initially focused on a subset of 54,000 models from Objaverse-XL provided by Diffusion4D~\cite{liang2024diffusion4d}, as these models exhibit movable characteristics and better geometric quality compared to the full dataset. However, many of these models were unsuitable for our purposes due to issues such as scene-based animations (multiple objects combined), the absence of skeletons or skinning weights, and a heavy bias towards human body-related models. This necessitated a rigorous preprocessing pipeline to create a high-quality dataset suitable for training our model.}

\subsubsection{Dataset Preprocessing}

Our preprocessing pipeline addressed the aforementioned challenges through a combination of empirical rules and the use of vision-language models (VLMs). This pipeline involved the following key steps:

\begin{itemize}
    \item[1] \cyp{\textbf{Skeleton-Based Filtering:}} We retained only the 3D assets with a bone count within the range of $[10, 256]$, while ensuring that each asset has a single, connected skeleton tree. \cyp{This step ensured that each model had a well-defined skeletal structure while removing overly simplistic or complex models and scenes containing multiple objects.}
    
    \item[2] \cyp{\textbf{Automated Categorization:}} We rendered each object under consistent texture and illumination conditions and deduplicated objects by computing the perceptual hashing value of the rendered images~\cite{farid2021overview}. \cyp{We then employed the vision-language model ChatGPT-4o~\cite{hurst2024gpt} to generate descriptive captions for each model. These captions were used to categorize the models into eight groups: Mixamo, Biped, Quadruped, Bird \& Flyer, Insect \& Arachnid, Water Creature, Static, and Other. Specifically, Static means some static objects such as pillows. This categorization, based on semantic understanding, allowed us to address the long-tail distribution problem and ensure sufficient representation of various object types.} Notably, we pre-screened skeletons conforming to the Mixamo~\cite{blackman2014rigging} format by their bone names and placed them in a separate category.
    
    \item[3] \cyp{\textbf{Manual Verification and Refinement:}} We re-rendered each model with its skeleton displayed to enable manual inspection of the skeletal structure and associated data. This crucial step allowed us to identify and correct common errors. One such issue is the incorrect marking of bone edges as ``not connected,'' which can result in many bones being directly connected to the root and an unreasonable topology. These issues introduce bias during network training and deviate from expected anatomical configurations. Specific corrections are detailed in Appendix~\ref{appendix:topo_mistake}.
\end{itemize}

\subsubsection{Dataset Details}

\begin{figure}[!t]
    \centering
    \includegraphics[width=\linewidth]{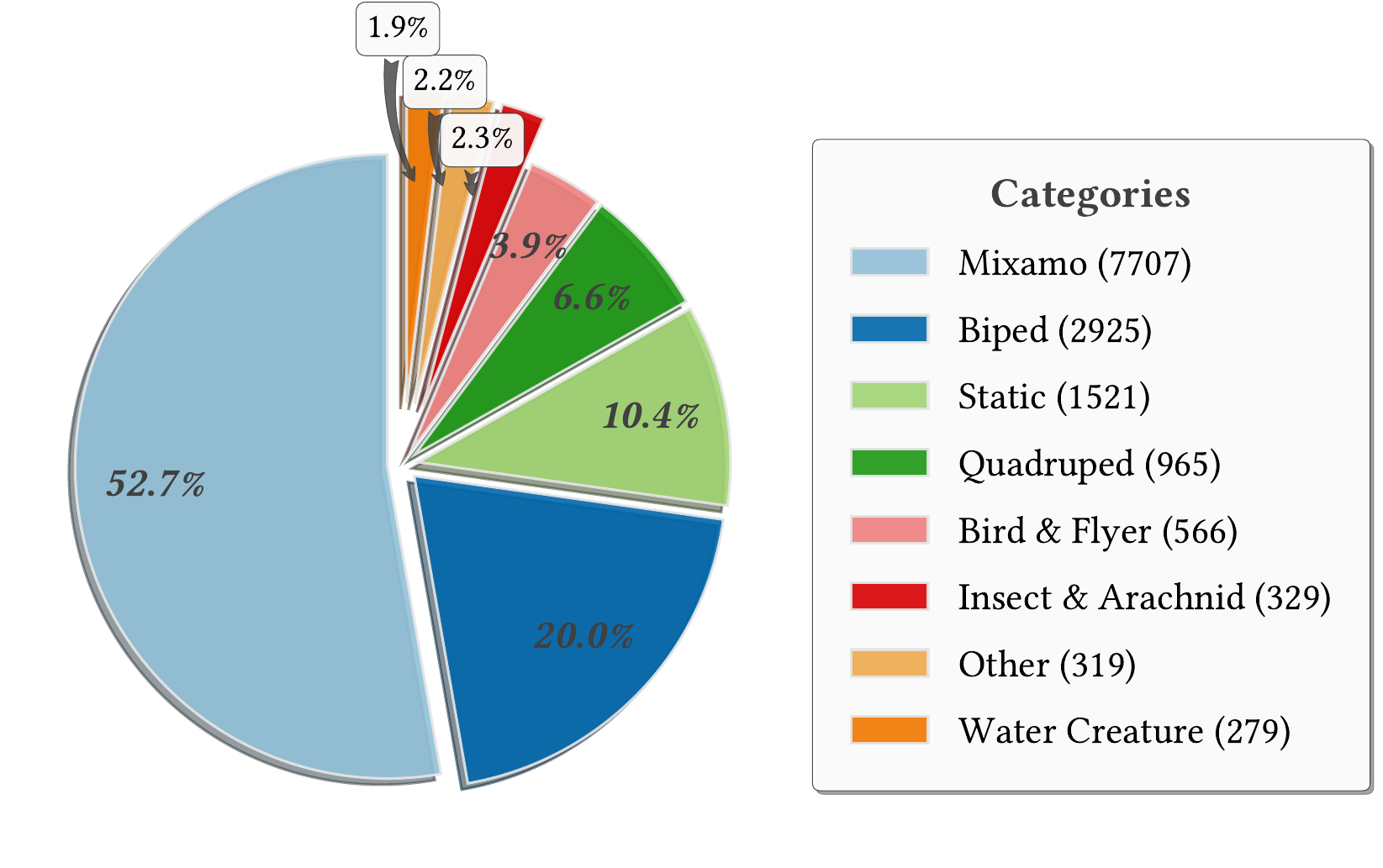}
    \caption{\textbf{Category distribution of \Dataset.} The percentages indicate the proportion of models belonging to each category.}
    \label{fig:proportion of the dataset}
\end{figure}

\begin{figure}[!t]
    \centering
    \includegraphics[width=\linewidth]{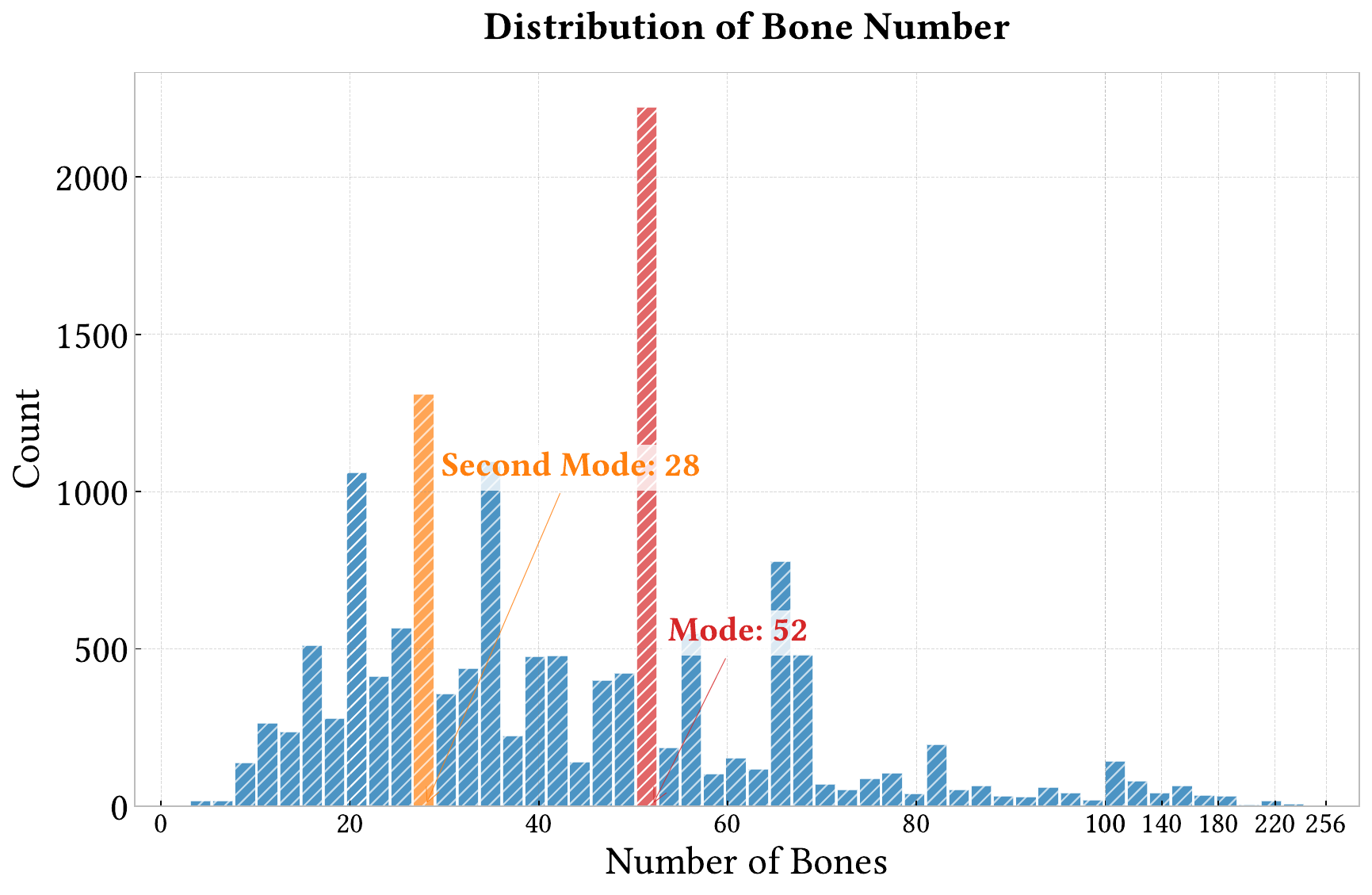}
    \caption{\textbf{Distribution of bone numbers in \Dataset.} The histogram shows the frequency of different bone counts across all models in the dataset.}
    \label{fig:skeleton amount of the dataset}
\end{figure}

\cyp{After this rigorous preprocessing, the \Dataset dataset comprises $14,611$ unique 3D models, each with a well-defined skeleton and complete skinning weights. The distribution across the eight categories is shown in~\ref{fig:proportion of the dataset}. Notably, human-related models (Mixamo and Biped) are still dominant, reflecting the composition of the original Objaverse-XL. \ref{fig:skeleton amount of the dataset} shows the distribution of skeleton counts, with a primary mode at $52$, corresponding to Mixamo models with hands, and a secondary mode at $28$, corresponding to Mixamo models without hands. This detailed breakdown of the dataset's composition highlights its diversity and suitability for training a generalizable rigging model.}

\begin{figure*}[!t]
    \centering
    \includegraphics[width=\linewidth]{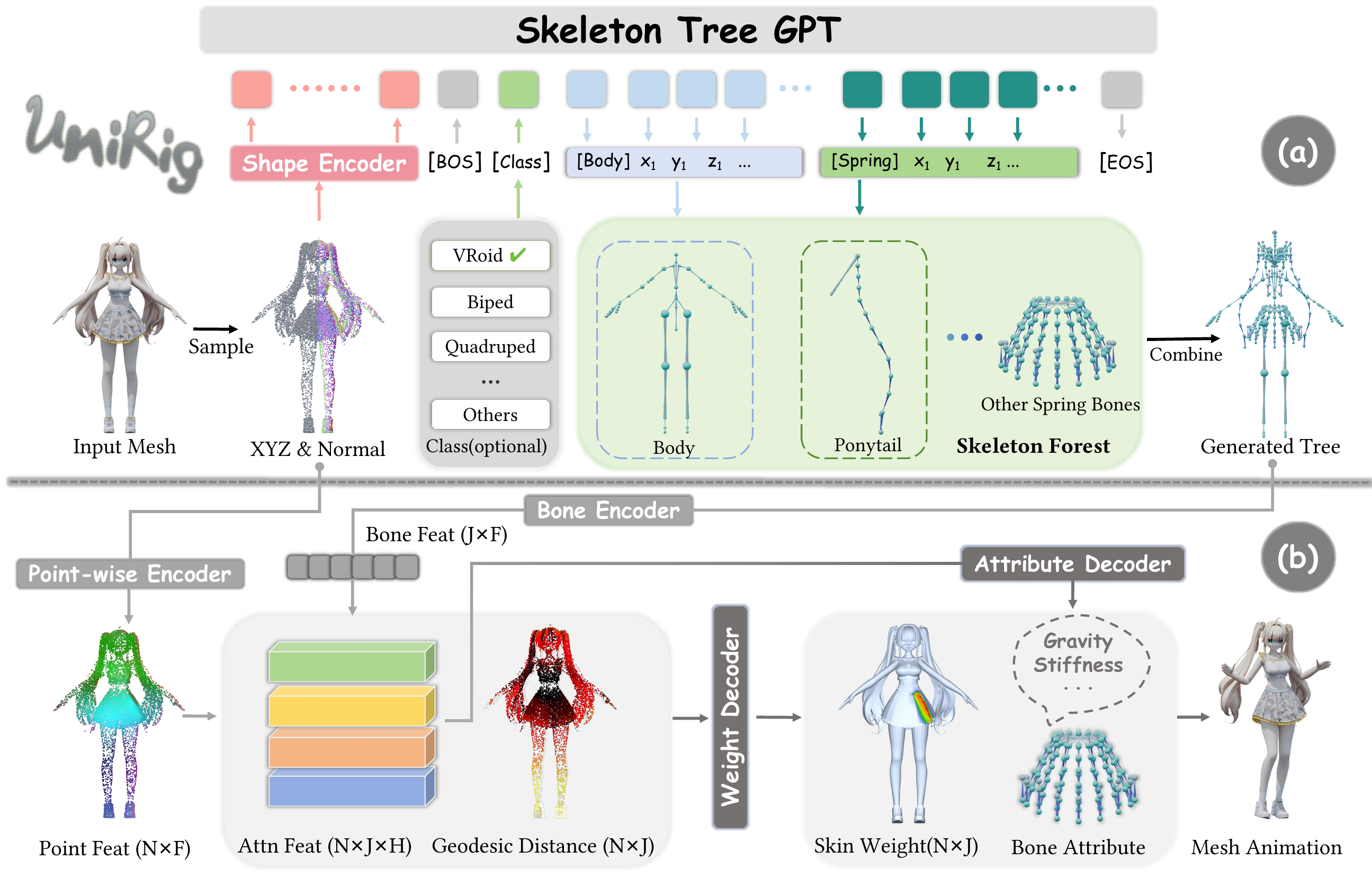}
    \caption{{\bf Overview of the \Method framework.} The framework consists of two main stages: (a) \textbf{Skeleton Tree Prediction} and (b) \textbf{Skin Weight Prediction}.  \textbf{(a)} The skeleton prediction stage (detailed in Section~\ref{sec:ar-skeleton}) takes a point cloud sampled from the 3D meshes as input, which is first processed by the Shape Encoder to extract geometric features. These features, along with optional class information, are then fed into an autoregressive Skeleton Tree GPT to generate a token sequence representing the skeleton tree. The token sequence is then decoded into a hierarchical skeleton structure. 
    \textbf{(b)} The skin weight prediction stage (detailed in Section~\ref{sec:skin_pred}) takes the predicted skeleton tree from (a) and the point cloud as input. A Point-wise Encoder extracts features from the point cloud, while a Bone Encoder processes the skeleton tree. These features are then combined using a Bone-Point Cross Attention mechanism to predict the skinning weights and bone attributes. Finally, the predicted rig can be used to animate the mesh. $\copyright$ kinoko7}
    \label{fig:method}
\end{figure*}

\section{Autoregressive Skeleton Tree Generation} \label{sec:ar-skeleton}

Predicting a valid and well-formed skeleton tree from a 3D mesh is a challenging problem due to the complex interdependencies between joints and the need to capture both the geometry and topology of the underlying structure. Unlike traditional methods that often rely on predefined templates or struggle with diverse topologies, we propose an autoregressive approach that generates the skeleton tree sequentially, conditioning each joint prediction on the previously generated ones. This allows us to effectively model the hierarchical relationships inherent in skeletal structures and generate diverse, topologically valid skeleton trees.

Formally, let \(\mathcal{M} = \{\mathcal{V} \in \mathbb{R}^{V \times 3}, \mathcal{F}\}\) represent a 3D mesh, where \(\mathcal{V}\) denotes the set of vertices and \(\mathcal{F}\) represents the faces. Our goal is to predict the joint positions \(\mathcal{J} \in \mathbb{R}^{J \times 3}\), where \(J\) is the number of bones, along with the joint-parent relationships \(\mathcal{P} \in \mathbb{N}^{J-1}\) that define the connectivity of the skeleton tree.

To facilitate this prediction, we first convert the input mesh ($\mathcal{M}$) into a point cloud representation that captures both local geometric details and overall shape information. We sample \(N = 65536\) points from the mesh surface \(\mathcal{F}\), yielding a point cloud \(\mathcal{X} \in \mathbb{R}^{N \times 3}\) and corresponding normal vectors \(\mathcal{N} \in \mathbb{R}^{N \times 3}\). Point clouds provide a flexible and efficient representation for capturing the geometric features of 3D shapes, and the inclusion of surface normals encodes important information about local surface orientation. The point cloud is normalized to coordinates within the range \([-1, 1]^3\). These vectors are then passed through a geometric encoder \(E_G: (\mathcal{X}, \mathcal{N}) \mapsto \mathcal{F}_G \in \mathbb{R}^{\zjp{V \times F}}\), \zjp{ where $F$ denotes the feature dimension,} generating the geometric embedding \(\mathcal{F}_G\). We utilize a shape encoder based on the 3DShape2Vecset representation~\cite{zhang20233dshape2vecset} due to its proven ability to capture fine-grained geometric details of 3D objects. For the encoder \(E_G\), we do not use any pretrained weights but instead initialize its parameters randomly using a Gaussian distribution. The resulting geometric embedding \(\mathcal{F}_G\) serves as a conditioning context for the autoregressive generation process.

We employ an autoregressive model based on the OPT architecture~\cite{zhang2022opt} to sequentially generate the skeleton tree. OPT's decoder-only transformer architecture is well-suited for this task due to its ability to model long-range dependencies and generate sequences in a causally consistent manner. To adapt OPT for skeleton tree generation, we first need to represent the tree \(\{\mathcal{J}, \mathcal{P}\}\) as a discrete sequence \(\mathcal{S}\). This is achieved through a novel tree tokenization process (detailed in Section~\ref{sec:tokenization}) that converts the tree structure into a sequence of tokens, enabling the autoregressive model to process it effectively.

During training, the autoregressive model is trained to predict the next token in the sequence based on the preceding tokens and the geometric embedding \(\mathcal{F}_G\). This is achieved using the Next Token Prediction (NTP) loss, which is particularly well-suited for training autoregressive models on sequential data. The NTP loss is formally defined as:
\[
\mathcal{L}_{\text{NTP}} = - \sum_{t=1}^{T} \log P(s_t \mid s_1, s_2, \dots, s_{t-1}, \mathcal{F}_G),
\]
where \(T\) denotes the total sequence length \(\mathcal{S} = \{s_1, s_2, \dots, s_T\}\), and \(P(s_t \mid s_1, \dots, s_{t-1})\) is the conditional probability of token \(s_t\) given the preceding tokens in the sequence. By minimizing this loss, the model learns to generate skeleton trees that are both geometrically consistent with the input mesh and topologically valid\zjp{, as evidenced by the quantitative results in Table~\ref{tab:bone_metric_j2j} and Supplementary Table~\ref{tab:bone_metric_jb2b}}. The geometric embedding \(\mathcal{F}_G\) is prepended to the tokenized sequence to provide the necessary geometric context for the autoregressive generation.

\subsection{Skeleton Tree Tokenization \label{sec:tokenization}}

A core challenge in autoregressively predicting skeleton trees is representing the tree structure in a sequential format suitable for a transformer-based model. This involves encoding both the spatial coordinates of each bone and the hierarchical relationships between bones. A naive approach would be to simply concatenate the coordinates of each bone in a depth-first or breadth-first order. However, this approach leads to several challenges, including difficulty in enforcing structural constraints, redundant tokens and inefficient training and inference.

To address these challenges, we propose a novel skeleton tree tokenization scheme. Inspired by recent advances in 3D generative model~\cite{chen2024meshanything, hao2024meshtron, siddiqui2024meshgpt}, our method discretizes the continuous bone coordinates and employs special tokens to represent structural information. \zjp{While inspired by these 3D generation approaches, our tokenization scheme is specifically designed for the unique challenge of representing the \textit{hierarchical structure} of a skeleton tree in a sequential format suitable for autoregressive rigging.}

We first discretize the normalized bone coordinates, which lie in the range \([-1, 1]\), into a set of \(D \zjp{=256}\) discrete tokens. This is done by mapping the continuous values to integers using the following function: \(M: x \in [-1, 1] \mapsto d = \lfloor \displaystyle \frac{x + 1}{2} \times D \rfloor \in \mathbb{Z}_D\). The inverse mapping is given by: \(M^{-1}: d \in \mathbb{Z}_D \mapsto x = \displaystyle \frac{2d}{D} - 1 \in [-1, 1]\). This discretization allows us to represent bone coordinates as sequences of discrete tokens. \zjp{The average relative error during discretization is $\mathcal O(\displaystyle \frac{1}{D})$, which is negligible for our application.}

Let \(\mathcal{J}_i\) be the $i$-th joint in the skeleton tree. We define the discrete index of the $i$-th bone as \(d_i = (dx_i, dy_i, dz_i)\), where \(dx_i = M(\mathcal{J}_i(x))\), \(dy_i = M(\mathcal{J}_i(y))\), and \(dz_i = M(\mathcal{J}_i(z))\) are the discretized coordinates of the tail of the $i$-th bone.

A straightforward way to tokenize the skeleton tree would be to concatenate these bone tokens in a topological order (e.g., depth-first), resulting in a sequence like:
\[
\begin{aligned}
\textbf{<bos>}~dx_1~dy_1~dz_1~dx_{\mathcal{P}_2}~dy_{\mathcal{P}_2}~dz_{\mathcal{P}_2}~dx_2~dy_2~dz_2 \cdots \\
dx_{\mathcal{P}_T}~dy_{\mathcal{P}_T}~dz_{\mathcal{P}_T}~dx_T~dy_T~dz_T~\textbf{<eos>}
\end{aligned}
\]
where \(\textbf{<bos>}\) and \(\textbf{<eos>}\) denote the beginning and end of the sequence, respectively, and $\mathcal{P}_i$ denotes the parent joint of the $i$-th joint.

However, this naive approach has several drawbacks. First, it introduces redundant tokens, as the coordinates of a joint are repeated for each of its children. Second, it does not explicitly encode the different types of bones (e.g., spring bones, template bones), which can have different structural properties. Finally, during inference, we observed that this representation often leads to repetitive token sequences.

To overcome these limitations, we propose an optimized tokenization scheme that leverages the specific characteristics of skeletal structures. Our key insight is that \zjp{decomposing skeleton tree into} certain bone sequences, such as spring bones in VRoid models or bones belonging to a known template (e.g., Mixamo), can be represented more compactly. Furthermore, explicitly encoding these bone types using dedicated type identifiers provides valuable information to the model, improving its ability to learn and generalize to different skeletal structures. For instance, knowing that a bone belongs to a specific template (e.g., Mixamo) allows for efficient motion retargeting, as the mapping between the template and the target skeleton is already known.

We introduce special ``type identifier'' tokens, denoted as \(\textbf{<type>}\), to indicate the type of a bone sequence. For example, a sequence of \zjp{spring bone chain} can be represented as 
\[
\begin{aligned}
\textbf{<spring\_bone>}~dx_s~dy_s~dz_s~...~dx_t~dy_t~dz_t, 
\end{aligned}
\]
where \(dx_s~dy_s~dz_s\) and \(dx_t~dy_t~dz_t\) are the discretized coordinates of the first and last spring bones in the \zjp{chain}, respectively. Similarly, bones belonging to a template can be represented using a template identifier, such as \(\textbf{<mixamo:body>}\). This allows us to omit the parent coordinates for bones in a template, as they can be inferred from the template definition. We also add a class token \(\textbf{<cls>}\) (e.g. \(\textbf{<mixamo>}\)) at the beginning of each sequence.

This results in a more compact tokenized sequence:
\[
\begin{aligned}
\textbf{<bos>}~\textbf{<cls>}~\textbf{<type}_1\textbf{>}~dx_1~dy_1~dz_1~dx_2~dy_2~dz_2 \cdots \textbf{<type}_2\textbf{>} \dots \\
\textbf{<type}_k\textbf{>} dx_t~dy_t~dz_t \dots dx_T~dy_T~dz_T~\textbf{<eos>}
\end{aligned}
\]

\begin{algorithm}[!h]
\caption{Skeleton Tree Tokenization \label{code:tokenize}}
\KwIn{$\textit{bones}$ $\cal B = (\cal J_{\cal P}, \cal J) \in \bb R^{J \times 6}$ (with skeleton Tree structure), $\textit{templates}$ $\cal T$ and class type of dataset $\cal C$}
\KwOut{$\textit{token sequence}$ $\cal S \in \bb N^T$}
\SetKwFunction{FMain}{tokenize}
\SetKwProg{Fn}{Function}{:}{}
\SetAlgoNlRelativeSize{0.5} 

\Fn{\FMain{$\textit{bones } \cal B, \textit{templates } \cal T, \textit{class type } \cal C$}}
{
    $d_i = (dx_i, dy_i, dz_i) \gets (M(\cal J_i (x)) M(\cal J_i(y)), M(\cal J_i(z)))$ \;
    $\cal S \gets  [\textbf{<bos>, <} \cal C \textbf{}{>}] $\;
    \texttt{Match Set} $ \cal M \gets \emptyset$; \tcp{Store the match bones}
    \For {
    $\texttt{template~} P \in \cal T$
    } {
        \If{$\cal B$ match $P$}{ \tcp{\zjp{$\cal B$ match $P$: requires tree structure and name matching}}
            $\cal S \gets  [\cal S, \textbf{<tempalte\_token of } P \textbf{>}]$ \;
            $\cal S \gets  [\cal S, dx_{P_0}, dy_{P_0}, dz_{P_0}, \dots, dx_{P_{|P|}}, dy_{P_{|P|}}, dz_{P_{|P|}}]$\;
            $\cal M \gets \{\cal M, P\}$
        }
    }

    \For{$R \in \cal J$}{
        \If{$R \not \in \cal M \textbf{ and } \cal P_R \in \cal M$}{ 
            \tcp{check $R$ is a root of remain forests}
            \texttt{stack.push($R$)}\;
            \texttt{last\_bone} $\gets$ \texttt{None}\;
            \While{$|\texttt{stack}| > 0$}{
                \texttt{bone} $b$ $\gets$ \texttt{stack.top()}; \tcp{get bone index $b$} 
                \texttt{stack.pop()}\;
                
                \If{$\texttt{parent}[b]$ $\neq$ \texttt{last\_bone}}{
                    $\cal S \gets [\cal S, \textbf{<branch\_token>}]$ \;
                    $\cal S \gets [\cal S, dx_{\cal P_b}, dy_{\cal P_b}, dz_{\cal P_b}]$ \;
                }
                $\cal S \gets [\cal S, dx_{b}, dy_{b}, dz_{b}]$ \;
                
                \texttt{last\_bone} $\gets$ $b$\;
                $\texttt{children}[b]$ \textbf{sorted by} $(z, y, x)$\;
                $\texttt{stack.push(children}[b]\texttt{)}$\;
            }
        }
    }
    $\cal S \gets [\cal S, \textbf{<eos>}]$\;
    \Return $\cal S$\;
}
\end{algorithm}

For more general cases where no specific bone type can be identified, we use a Depth-First Search (DFS) algorithm to identify and extract linear bone chains, and represent them as compact subsequences. \zjp{The DFS traversal identifies separate bone chains (branches) originating from the main skeleton structure or forming disconnected components. Each newly identified branch is then prefixed with a \textbf{<branch\_token>} in the token sequence.}  We also ensure the children of each joint are sorted based on their tail coordinates \zjp{\((z, y, x)\) order} in the rest pose\zjp{ (where the $z$-axis represents the vertical direction in our coordinate convention)}. This maintains a consistent ordering that respects the topological structure of the skeleton. The specific steps of this optimized tokenization process are summarized in Algorithm~\ref{code:tokenize}.

For instance, consider an anime-style 3D girl with a spring-bone-based skirt, as shown in Figure~\ref{fig:method}(a). Using our optimized tokenization, this could be represented as:
\[
\begin{aligned}
\textbf{<bos>}~\textbf{<VRoid>}~\textbf{<mixamo:body>}~dx_1~dy_1~dz_1 \dots dx_{22}~dy_{22}~dz_{22} \\
\textbf{<mixamo:hand>}~dx_{23}~dy_{23}~dz_{23} \dots dx_{52}~dy_{52}~dz_{52} \dots \\
\textbf{<spring\_bone>}~dx_s~dy_s~dz_s \dots dx_t~dy_t~dz_t \dots \textbf{<eos>}
\end{aligned}
\]
This demonstrates how our tokenization scheme compactly represents different bone types and structures. 

\zjp{During de-tokenization, connectivity between different bone chains (identified by their respective tokens) is established by merging joints whose decoded coordinates fall within a predefined distance threshold, effectively reconstructing the complete skeleton tree.}

This optimized tokenization significantly reduces the sequence length compared to the naive approach. Formally, the naive approach requires \(6T - 3 + K\) tokens (excluding \(\textbf{<bos>}\) and \(\textbf{<eos>}\)), where $T$ is the number of bones. In contrast, our optimized tokenization requires only \(3T + M + S \times 4 + 1\) tokens, where $M$ is the number of templates (usually less than $2$), and $S$ is the number of branches in the skeleton tree after removing the templates to form a forest. As shown in Table~\ref{tab:token_cost_compare}, we observe an average token reduction of $27.47\%$ on VRoid and $29.72\%$ on \Dataset. 

In addition to reducing the number of tokens required to represent the skeletal tree, our representation ensures that when generating based on a template, the generated fixed positions correspond precisely to the skeleton. By leveraging positional encoding and an autoregressive model, this tokenization approach enables higher accuracy in template-specified predictions.
These lead to reduced memory consumption during training and faster inference, making our method more efficient.

\begin{table}[!t]
    \centering
    \caption{The average token costs in representing a skeleton tree of different datasets. Our optimized tokenization can reduce about $30 \%$ tokens.}
    \begin{tabular}{c|c|c|c}
        \hline
        \diagbox{Dataset}{Method} &  Naïve & Optimized & Tokens Reduction\\
        \hline
        VRoid & 667.27 & 483.95 & 27.47 \% \\
        \Dataset & 266.28 & 187.15 & 29.72 \% \\
        \hline
    \end{tabular}
    \label{tab:token_cost_compare}
\end{table}

\section{Skin Weight Prediction via Bone-Point Cross Attention\label{sec:skin_pred}}

Having predicted the skeleton tree in Section 5, we now focus on predicting the skinning weights that govern mesh deformation. These weights determine the influence of each bone on each vertex of the mesh. Formally, we aim to predict a weight matrix $\mathcal{W} \in \mathbb{R}^{N \times J}$, where $N$ is the number of vertices in the mesh and $J$ is the number of bones. In our case, $N$ can be in the tens of thousands due to the complexity of models in \Dataset, and $J$ can be in the hundreds. The high dimensionality of $\mathcal{W}$ poses a significant computational challenge.

Additionally, many applications require the prediction of bone-specific attributes, denoted by $\mathcal{A} \in \mathbb{R}^{J \times B}$, where $B$ is the dimensionality of the attribute vector. These attributes can encode various physical properties, such as stiffness or gravity coefficients, which are crucial for realistic physical simulations (detailed in Section~\ref{sec:physics}). Some bones might also act purely as connectors without influencing mesh deformation, as indicated by the ``connected'' option in Blender~\cite{blender}.

To address these challenges, we propose a novel framework for skin weight and bone attribute prediction that leverages a bone-informed cross-attention mechanism~\cite{vaswani2017attention}. This approach allows us to efficiently model the complex relationships between the predicted skeleton and the input mesh.

Our framework utilizes two specialized encoders: a bone encoder \( E_B \) and a point-wise encoder \( E_P \). The bone encoder, \( E_B \), is a Multi-Layer Perceptron (MLP) with positional encoding that processes the head and tail coordinates of each bone, represented as $(\mathcal{J}_{\mathcal{P}}, \mathcal{J}) \in \mathbb{R}^{J \times 6}$. This yields bone features \( \mathcal{F}_B \in \mathbb{R}^{J \times F} \), where $F$ is the feature dimensionality.

For geometric feature extraction, we employ a pretrained Point Transformer V3~\cite{wu2024pointv3} as our point-wise encoder, \( E_P \). Specifically, we use the architecture and weights from SAMPart3D~\cite{yang2024sampart3d}, which was pretrained on a large dataset of 3D objects~\cite{deitke2024objaverse}. SAMPart3D's removal of standard downsampling layers enhances its ability to capture fine-grained geometric details. The point-wise encoder takes the input point cloud, $\mathcal{X} \in \mathbb{R}^{N \times 3}$, and produces point-wise features $\mathcal{F}_P \in \mathbb{R}^{N \times F}$.

To predict skinning weights, we incorporate a cross-attention mechanism to model the interactions between bone features and point-wise features. We project the point-wise features \( \mathcal{F}_P \) into query vectors \( \mathcal{Q}_W \), and the bone features \( \mathcal{F}_B \) to key and value vectors \( \mathcal{K}_W \) and \( \mathcal{V}_W \). The attention weights \( \mathcal{F}_W \in \mathbb{R}^{N \times J \times H} \) are then computed as:
\[
\mathcal{F}_W = \text{softmax} \left( \frac{\mathcal{Q}_W \mathcal{K}_W^T}{\sqrt{F}} \right),
\]
where \( H \) is the number of attention heads. Each element \( \mathcal{F}_W(i, j) \) represents the attention weight between the $i$-th vertex and the $j$-th bone, essentially capturing the influence of each bone on each vertex.

We further augment the attention weights by incorporating the \zjp{voxel }geodesic distance\zjp{\cite{dionne2013geodesic}} \( \mathcal{D} \in \mathbb{R}^{N \times J} \) between each vertex and each bone, following previous work~\cite{xu2022morig, xu2020rignet}. This distance provides valuable information about the spatial proximity of bones and vertices, which is crucial for accurate skin weight prediction. The geodesic distance $\mathcal{D}$ is precomputed and concatenated with the attention weights \( \mathcal{F}_W \). Finally, the skinning weights \( \mathcal{W} \) are obtained by passing the concatenated features through an MLP, \( E_W \), followed by a softmax layer for normalization:
\[
\mathcal{W} = \text{softmax} \left( E_W \left( \text{concat} \left( \text{softmax} \left( \frac{\mathcal{Q}_W \mathcal{K}_W^T}{\sqrt{F}} \right), \mathcal{D} \right) \right) \right).
\]

For the prediction of bone attributes $\mathcal{A}$, we reverse the roles of bones and vertices in the cross-attention mechanism. Bone features $\mathcal{F}_B$ become the query, and point-wise features $\mathcal{F}_P$ are projected to key and value vectors. The bone attributes are then predicted using another MLP, \( E_A \):
\[
\mathcal{A} = E_A \left( \text{cross\_attn} \left( \mathcal{F}_B, \mathcal{F}_P \right) \right).
\]
We use the Kullback-Leibler (KL) divergence~\cite{van2014renyi} between the predicted and ground-truth skinning weights (\( \mathcal{W}_{\text{pred}} \) and \( \mathcal{W} \)) and the L2 loss between the predicted and ground-truth bone attributes (\( \mathcal{A}_{\text{pred}} \) and \( \mathcal{A} \)). The combined loss function is given by:
\[
\lambda_{\mathcal{W}} \mathcal{L}_{\text{KL}}(\mathcal{W}, \mathcal{W}_{\text{pred}}) + \lambda_{\mathcal{A}} \mathcal{L}_2 (\mathcal{A}, \mathcal{A}_{\text{pred}})
\]

\subsection{Training Strategy Based on Skeletal Equivalence}

A naive approach to training would involve uniformly sampling points from the mesh surface. However, this leads to an imbalance in the training of different bones. Bones in densely sampled regions, such as the hip, tend to learn faster than those in sparsely sampled regions, such as hair or fingers. Additionally, using hierarchical point cloud sampling based on skinning weights can introduce discrepancies between the training and inference processes, ultimately hurting the model's performance during inference.

To address these issues, we propose a training strategy based on \emph{skeletal equivalence}. Our key insight is that each bone should contribute equally to the overall training objective, regardless of the number of mesh vertices it influences. To achieve this, we introduce two key modifications to our training procedure. \emph{First}, during each training iteration, we randomly freeze a subset of bones with a probability \( p \). For these frozen bones, we use the ground-truth skinning weights and do not compute gradients. This ensures that all bones, even those in sparsely sampled regions, have an equal chance of being updated during training. \emph{Second}, we introduce a bone-centric loss normalization scheme. Instead of averaging the loss over all vertices, we normalize the loss for each bone by the number of vertices it influences. This prevents bones that influence many vertices from dominating the loss function. Formally, our normalized loss function is given by:
\[
\sum_{i=1}^J \frac{1}{J} \sum_{k=1}^N \frac{[\mathcal{W}_{k,i} > 0] \mathcal{L}_2^{(k)}}{S_k = \sum_{k=1 \dots N} [\mathcal{W}_{k,i} > 0]} = \frac{1}{J} \sum_{k=1}^N \mathcal{L}_2^{(k)} \left( \sum_{i=1}^J \frac{[\mathcal{W}_{k,i} > 0]}{S_k} \right),
\]
where \( S_k \) denotes the normalization factor based on the number of active points in each bone. It means we average the loss weight according to bone number instead of sample point number.
where $J$ is the number of bones, $N$ is the number of vertices, and $[\mathcal{W}_{k,i} > 0]$ is an indicator function(iverson bracket) that is $1$ if vertex $i$ is influenced by bone $j$, and $0$ otherwise. This can also be interpreted as first averaging the loss for each bone, and then averaging across all bones. $\cal L_2^{(k)}$ means the $k$-th vertex reconstruction loss of indirect supervision in Section \ref{sec:physics}.
By incorporating these two techniques, our training strategy ensures that all bones are trained equally, leading to improved performance, especially for bones in sparsely sampled regions.

\subsection{Indirect Supervision via Physical Simulation\label{sec:physics}}
While direct supervision using skinning weight loss can yield good results, it may not always guarantee visually realistic motion. This is because different combinations of skinning weights can produce similar deformations under simple transformations, even if one set of weights is physically implausible. To address this issue, we introduce an indirect supervision method that incorporates physical simulation to guide the learning process toward more realistic results. This method provides a more robust training signal by evaluating the quality of the predicted skinning weights and bone attributes based on the resulting motion.

\begin{figure}[!t]
    \centering
    \includegraphics[width=0.8\linewidth]{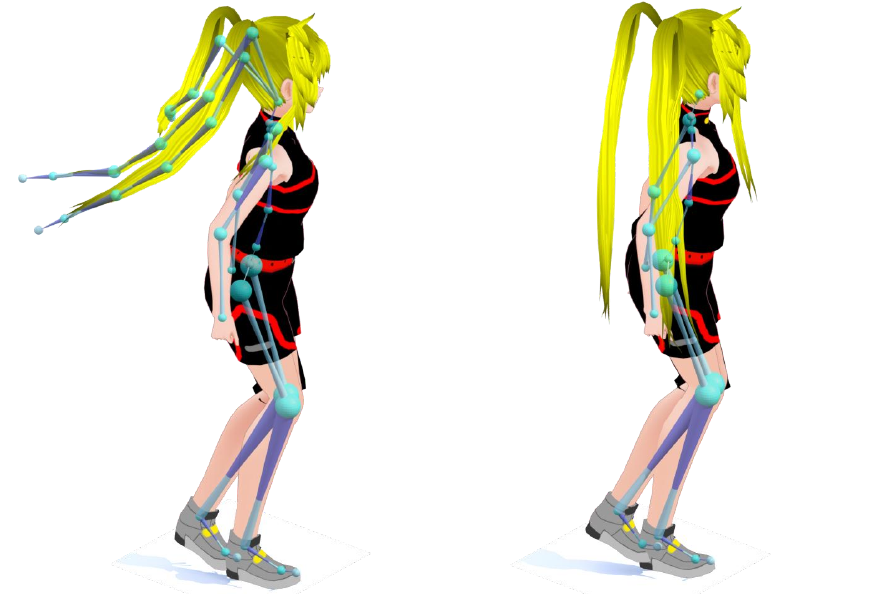}
    \caption{\textbf{Comparison of model animation with and without spring bones.} The model on the left utilizes spring bones, resulting in more natural and dynamic movement of the hair and skirt. The model on the right does not use spring bones, leading to a stiffer and less realistic appearance, with only rigid body motion.}
    \label{fig:compare_spring}
\end{figure}

Our approach extends beyond traditional Linear Blend Skinning (LBS) by incorporating a differentiable Verlet integration-based physical simulation, inspired by the spring bone dynamics in VRoid models~\cite{isozaki2021vroid}. This simulation allows us to model the behavior of bones under the influence of physical forces like gravity and stiffness, as defined by the predicted bone attributes. By comparing the simulated motion generated using the predicted parameters with that generated using the ground-truth parameters, we can obtain a more accurate measure of the prediction quality. Figure~\ref{fig:compare_spring} illustrates the impact of spring bones on the realism of the animation.

In the VRM standard, spring motion is governed by several physical parameters, including drag coefficient \( \eta_d \), stiffness coefficient \( \eta_s \), gravity coefficient \( \eta_g \), and gravity direction \( \mathbf{g} \). For simplicity, we assume a uniform downward gravity direction and neglect collisions. Verlet integration is used to compute the bone's tail position at each time step, requiring both the current and previous frames' positions. To prevent numerical instability, the bone length is normalized after each integration step. The details of the simulation are provided in Algorithm~\ref{code:physics} in the supplementary material.

To incorporate this physical simulation into our training, we randomly sample a short motion sequence $M$ \zjp{from the Mixamo dataset} of length $T$ and apply it to both the predicted and ground-truth parameters. This results in two sets of simulated vertex positions: \( \mathcal{X}^{\mathcal{M}}_{\text{pred}} \) (using predicted skinning weights $\mathcal{W}_{\text{pred}}$ and bone attributes $\mathcal{A}_{\text{pred}} \}$) and \( \mathcal{X}^{\mathcal{M}} \) (using ground-truth $\mathcal{W}$ and $\mathcal{A}$). To ensure gradient stability, we use a short sequence length of \( T = 3 \), which is sufficient to capture the effects of the physical simulation.

We then use the L2 distance between the simulated vertex positions as a reconstruction loss, which serves as our indirect supervision signal. This loss, combined with the direct supervision losses from Section~\ref{sec:skin_pred} forms our final loss function:
\[
\lambda_{\mathcal{W}} \mathcal{L}_{\text{KL}}(\mathcal{W}, \mathcal{W}_{\text{pred}}) + \lambda_{\mathcal{A}} \mathcal{L}_2 (\mathcal{A}, \mathcal{A}_{\text{pred}}) + \lambda_{\mathcal{X}} \sum_{i = 1}^T \mathcal{L}_2(\mathcal{X}^{\mathcal{M}_i}, \mathcal{X}^{\mathcal{M}_i}_{\text{pred}}).
\]
where $\lambda_{\mathcal{W}}$, $\lambda_{\mathcal{A}}$, and $\lambda_{\mathcal{X}}$ are weighting factors that balance the different loss terms. This combined loss function encourages the model to predict skinning weights and bone attributes that not only match the ground truth directly but also produce physically realistic motion.

\section{Experiments\label{sec:experiments}}

\subsection{Implementation Details}

\subsubsection{Dataset Preprocessing}

As illustrated in Figure \ref{fig:proportion of the dataset}, the original \Dataset dataset exhibits a highly skewed distribution, with human-related categories (Mixamo and Biped) being significantly overrepresented. Directly training on this unbalanced distribution would lead to suboptimal performance, particularly for underrepresented categories. To mitigate this issue and ensure a more balanced training set across diverse skeleton types, we adjusted the sampling probabilities for each category as follows: VRoid: $25\%$, Mixamo: $5\%$, Biped: $10\%$, Quadruped: $20\%$, Bird \& Flyer: $15\%$, Static: $5\%$, and Insect \& Arachnid: $10\%$. This distribution prioritizes high-quality data (VRoid) while ensuring sufficient representation of other categories.

To further enhance the robustness and generalizability of our model, we employed two key data augmentation techniques:
\begin{itemize}
    \item[1] \textbf{Random Rotation \& Scaling:} With a probability of $p_r=0.4$, we randomly rotated the entire point cloud around each of the three coordinate axes by an \zjp{Euler} angle $r \in [-30^\circ, 30^\circ]$ \zjp{(XYZ order)}. Independently, with a probability of $p_s=0.5$, we scaled the point cloud by a factor $s \in [0.8, 1.0]$.
    \item[2] \textbf{Motion-Based Augmentation:} We applied motion sequences to the models to augment the training data with a wider range of poses. For models in the Mixamo and VRoid categories, we applied motion sequences from the Mixamo action database with a probability of $p_{m1}=0.6$. For models in other categories, we randomly rotated individual bones with a probability of $p_{m2}=0.4$, with rotation angles sampled from $r \in [-15^\circ, 15^\circ]$.
\end{itemize}

\subsubsection{Training Strategy}

Our training process consists of two stages: skeleton tree prediction and skin weight prediction.
For \emph{skeleton tree prediction} (Section \ref{sec:ar-skeleton}), we employed the OPT-125M transformer \cite{zhang2022opt} as our autoregressive model, combined with a geometric encoder based on the 3DShape2Vecset framework \cite{zhang20233dshape2vecset, zhao2024michelangelo}. The model was trained for 3 days on 8 NVIDIA A100 GPUs, utilizing the AdamW optimizer \cite{loshchilov2017decoupled} with parameters $\beta_1 = 0.9$, $\beta_2 = 0.999$, and a weight decay of 0.01. We trained for a total of 500 epochs with a cosine annealing learning rate schedule, starting at a learning rate of  $1 \times 10^{-3}$ and decreasing to $2 \times 10^{-4}$. For \emph{skin weight prediction} (Section \ref{sec:skin_pred}), we sampled 16,384 points from each mesh during training. We used a reduced model to save training resources, which includes a frozen pretrained Point Transformer from SAMPart3D \cite{yang2024sampart3d} and only a small portion of parameters in the Bone Encoder, Cross Attention, and Weight Decoder modules are trainable. The learning rate was fixed at $1 \times 10^{-3}$ during this stage. This phase of training required 1 day on 8 NVIDIA A100 GPUs.

\subsection{Results and Comparison}

To evaluate the effectiveness of our proposed method, we conducted a comprehensive comparison against both state-of-the-art academic methods and widely used commercial tools. Our evaluation focuses on two key aspects: \emph{skeleton prediction accuracy} and \emph{skinning quality}.
For \emph{quantitative evaluation} of skeleton prediction, we compared UniRig with several prominent open-source methods: RigNet \cite{xu2020rignet}, NBS \cite{li2021learning}, and TA-Rig \cite{ma2023tarig}. These methods represent the current state-of-the-art in data-driven rigging. We used a validation set consisting of $50$ samples from the VRoid dataset and $100$ samples from the \Dataset dataset. The validation set and training dataset are guaranteed to never overlap after we deduplicate them carefully in Section \ref{sec:dataset-rigxl}. The validation samples in \Dataset are selected uniformly from each class. The VRoid samples allowed us to assess the performance on detailed, anime-style characters, while the \Dataset samples tested the generalizability of our method across diverse object categories.
We also performed a \emph{qualitative comparison} against several commercial and closed-source systems, including Meshy \cite{meshy}, Anything World \cite{anythingworld}, and Accurig \cite{accurig}. Due to the closed-source nature of these systems, a direct quantitative comparison was not feasible. Instead, we compared the visual quality of the generated skeletons and the resulting mesh animations. The qualitative results are presented and discussed.

\subsubsection{Bone Prediction\label{sec:bone_metric}}

To evaluate the accuracy of our bone prediction, we used three metrics based on chamfer distance:
\begin{itemize}
    \item \textbf{Joint-to-Joint Chamfer Distance (J2J):} Measures the average chamfer distance between corresponding predicted and ground-truth joint positions.
    \item \textbf{Joint-to-Bone Chamfer Distance (J2B):} Measures the average chamfer distance between predicted joint positions and their closest points on the ground-truth bone segments.
    \item \textbf{Bone-to-Bone Chamfer Distance (B2B):} Measures the average chamfer distance between points on the predicted bone segments and their closest points on the ground-truth bone segments.
\end{itemize}
Lower values for these metrics indicate better prediction accuracy. For a fair comparison with prior work on the Mixamo and VRoid datasets, we evaluated the metrics using a reduced set of 52 bones (or 22 bones). For the \Dataset dataset, which contains more diverse skeletal structures, we used the complete set of predicted bones. All mesh models were normalized to a unit cube ($[-1, 1]^3$) to ensure consistent evaluation across datasets. All mesh models were normalized to a unit cube ($[-1, 1]^3$) to ensure consistent evaluation across datasets.

Table \ref{tab:bone_metric_j2j} presents the quantitative results for the J2J metric. Our method, UniRig, outperforms all other methods across all datasets, demonstrating its superior accuracy in predicting joint positions. Additional results for the J2B and B2B metrics are provided in Supplementary Table \ref{tab:bone_metric_jb2b}, further demonstrating the effectiveness of our approach.

\begin{figure}[!t]
    \centering
    \includegraphics[width=\linewidth]{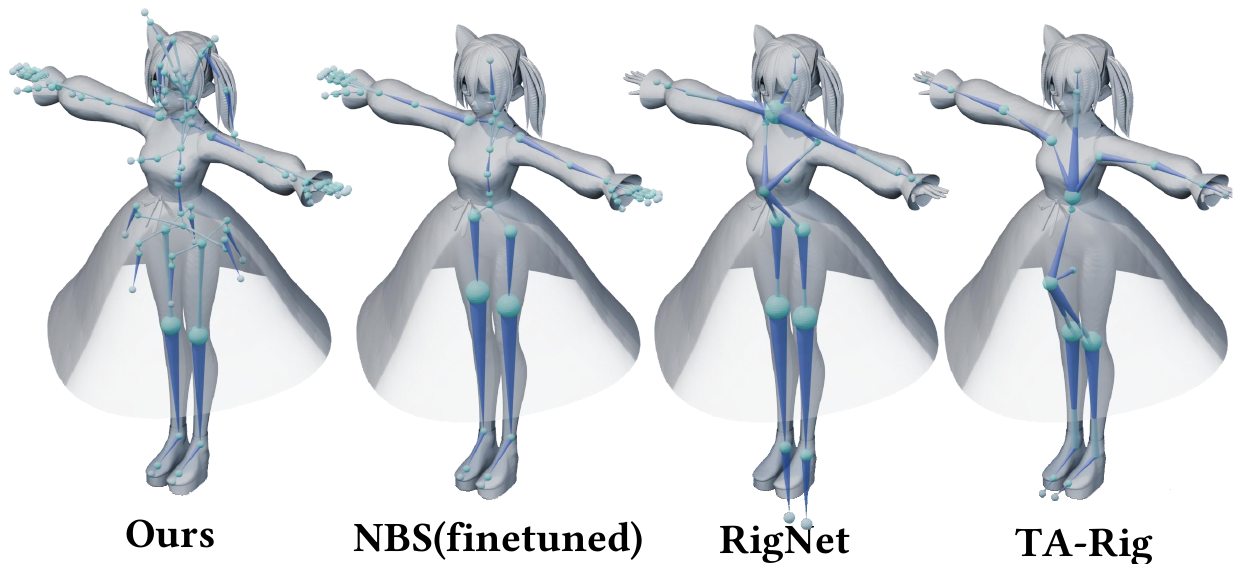}
    \caption{Comparison of predicted skeletons between NBS (fine-tuned), RigNet, and TA-Rig on the VRoid dataset. Our method (UniRig) generates skeletons that are more detailed and accurate.}
    \label{fig:compare_bone_pre}
\end{figure}

Figure \ref{fig:compare_bone_pre} provides a visual comparison of the predicted skeletons against RigNet, NBS, and TA-Rig on the VRoid dataset. The results show that UniRig generates more detailed and accurate skeletons. Further visual comparisons with academic methods are available in Supplementary Figure \ref{fig:compare_skeleton_academic}.

\begin{table}[!t]
 \renewcommand{\tabcolsep}{0.14mm}
    \centering
    \caption{Quantitative comparison of Joint-to-Joint Chamfer Distance (J2J).  $^{\ast}$ indicates the evaluation dataset is under the data augmentation of random rotation, scale, and applying random motion. $^\dagger$ indicates the model cannot be finetuned because RigNet does not provide data preprocess tools and TA-Rig does not provide training scripts. The best results are \textbf{bold}}
    \resizebox{1.05\linewidth}{!}{
    \begin{tabular}{l|c|c|c|c|c}
    \hline
     \diagbox[width=3.9cm]{Method}{Dataset}  & Mixamo & VRoid & Mixamo$^{\ast}$ & VRoid$^{\ast}$ & $\Dataset^{\ast}$ \\
         \hline
      Ours  & \bf{0.0101} & \bf{0.0092} & \bf{0.0103} & \bf{0.0101} & \bf{0.0549}\\
      RigNet$^\dagger$ \cite{xu2020rignet} & 0.1022  & 0.2405  & 0.2171 & 0.2484 & 0.2388\\
      NBS \cite{li2021learning} & 0.0338  & 0.0205 & 0.0429 & 0.0214 & N/A \\
      TA-Rig$^\dagger$ \cite{ma2023tarig}  & 0.1007  & 0.0886 & 0.1093 & 0.0934 & 0.2175\\
      \hline
    \end{tabular}
    }
    \label{tab:bone_metric_j2j}
\end{table}

\begin{figure}[!t]
    \centering
    \includegraphics[width=1.05\linewidth]{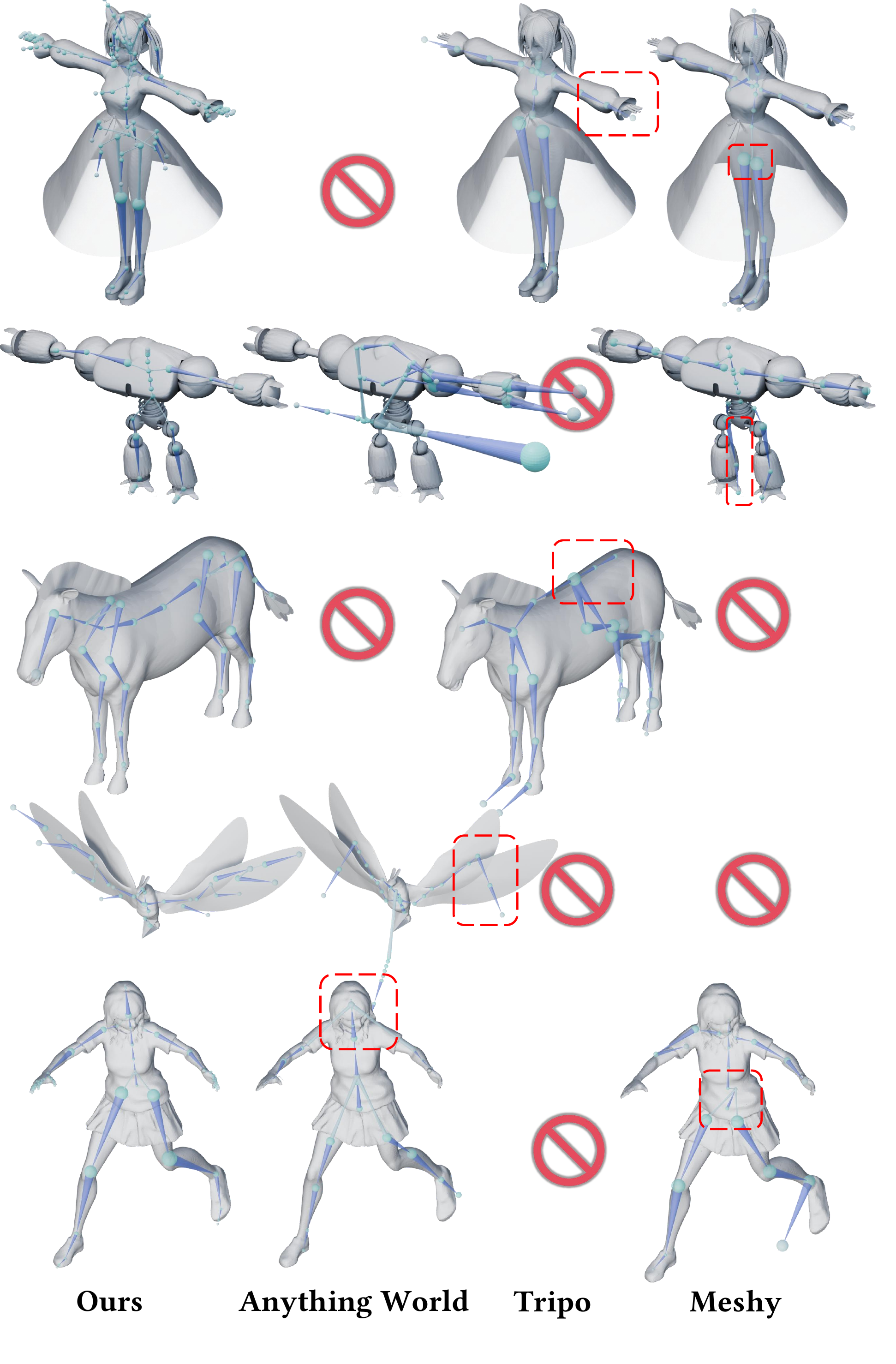}
    \vspace{-5mm}
     \caption{\textbf{Qualitative comparison of predicted skeletons against commercial tools.} Our method (UniRig) outperforms Tripo \cite{tripoai}, Meshy \cite{meshy}, Anything World \cite{anythingworld}, and Accurig \cite{accurig} in terms of both accuracy and detail. Red stop signs indicate that the corresponding tool failed to generate a skeleton.}
    \label{fig:compare_skeleton_commercial}
    \vspace{-5mm}
\end{figure}

We also conducted a qualitative comparison against commercial tools, including Tripo \cite{tripoai}, Meshy \cite{meshy}, and Anything World \cite{anythingworld}. As illustrated in Figure \ref{fig:compare_skeleton_commercial}, our method substantially outperforms these commercial systems, offering superior accuracy across a diverse range of mesh types, while also improving the completeness of the predicted skeletons.

\begin{table}[!t]
 \renewcommand{\tabcolsep}{0.14mm}
    \centering
    \caption{Comparison of skinning weight prediction accuracy using per-vertex L1 loss between predicted and ground-truth skinning weights. $^{\ast}$ means the evaluation dataset is under the data augmentation of random rotation, scale, and applying random motion. $^\dagger$ indicates the model cannot be finetuned because RigNet does not provide data preprocess tools and TA-Rig does not provide training scripts.}
    \resizebox{\linewidth}{!}{
    \begin{tabular}{l|c|c|c|c|c}
    \hline
      \diagbox[width=3.1cm]{Method}{Dataset}   & Mixamo & VRoid & Mixamo$^{\ast}$ & VRoid$^{\ast}$ & $\Dataset^{\ast}$ \\
         \hline
      Ours  & \bf{0.0055} & \bf{0.0028} & \bf{0.0059}& \bf{0.0038} & 0.0329 \\
      RigNet$^\dagger$ \cite{xu2020rignet} & 0.04540  & 0.04893  & 0.05367 & 0.06146 & N/A \\
      NBS\cite{li2021learning} & 0.07898  & 0.02721 & 0.08211 & 0.03339 & N/A \\
      \hline
    \end{tabular}
    }
    \label{tab:skin_metric}
    \vspace{-2mm}
\end{table}

\begin{table*}[!t]
    \centering
    \caption{Comparison of mesh deformation robustness using reconstruction loss under various animation sequences.  $^{\ast}$ means the evaluation dataset is under the data augmentation of random rotation, scale, and applying random motion. }
    \begin{tabular}{l|c|c|c|c|c|c}
    \hline
    \diagbox{Method}{Dataset}  & Mixamo & VRoid & Mixamo$^{\ast}$ & VRoid$^{\ast}$ & VRoid with Spring$^{\ast}$ & \Dataset \\
         \hline
      Ours  & $\bf 4.00 \times 10^{-4}$ & $\bf 4.00 \times 10^{-4}$ & $\bf 6.00 \times 10^{-4}$ & $\bf 1.10 \times 10^{-3}$ & $\bf 1.70 \times 10^{-3}$ & $3.5 \times 10^{-3}$\\
      NBS \cite{li2021learning} & $8.03 \times 10^{-4}$ & $5.82 \times 10^{-2}$ & $1.38 \times 10^{-3}$ & $2.34 \times 10^{-3}$ & $2.71 \times 10^{-3}$ & N/A \\
      \hline
    \end{tabular}
    \label{tab:motion_metric}
\end{table*}

\begin{figure*}[!t]
    \centering
    \includegraphics[width=0.95\linewidth]{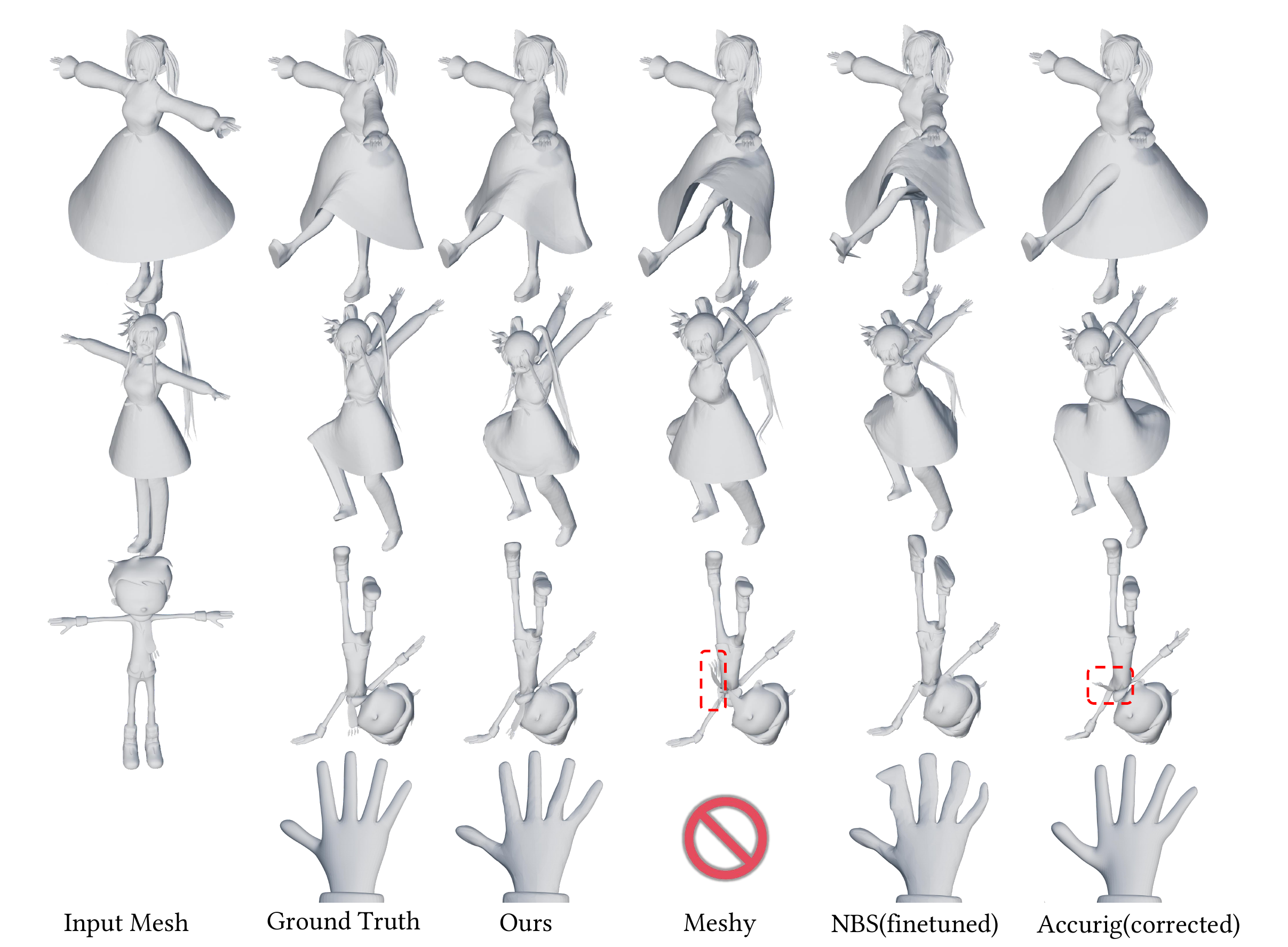}
    \caption{\textbf{Qualitative comparison of mesh deformation under motion.} Our method (UniRig) is compared with commercial tools (Meshy \cite{meshy} and Accurig \cite{accurig}) and a state-of-the-art academic method (NBS \cite{li2021learning}) on several models. Our model and the ground truth both exhibit realistic physical simulation of spring bones, resulting in more natural hair and clothing movement. Our method also demonstrates precise hand weight prediction, enabling fine-grained hand movements. Note that NBS was fine-tuned on the VRoid dataset, while Accurig requires joint manually corrected.}
    \label{fig:compare_motion_commercial}
\end{figure*}

\begin{figure*}[!t]
    \centering
    \includegraphics[width=0.9\linewidth]{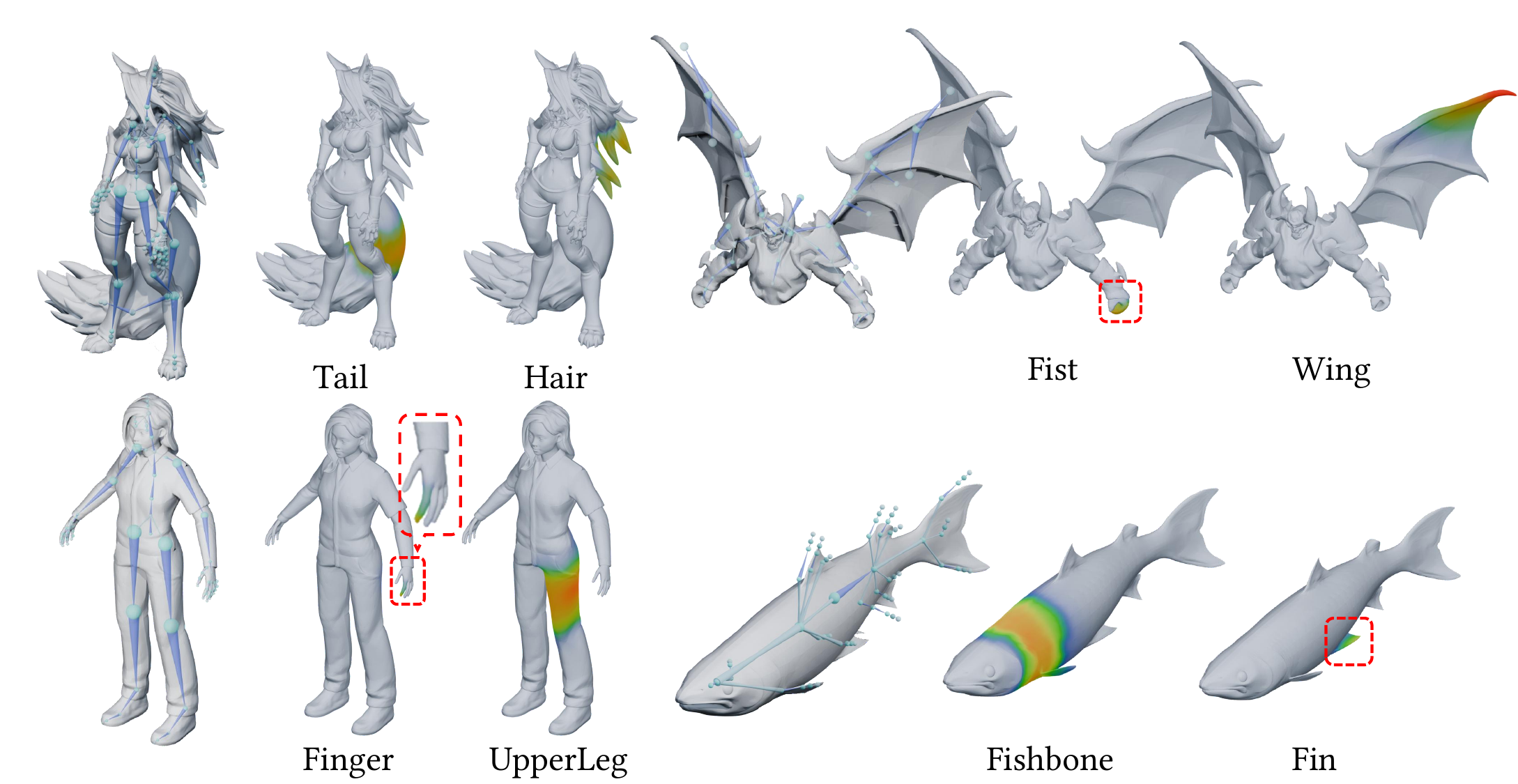}
    \caption{\textbf{Qualitative results of \Method on various object categories.} The figure showcases the predicted skeletons, skinning weights, and the resulting deformed meshes. Our method demonstrates the ability to predict highly detailed skeletal structures and accurate local skin weight mappings.}
    \label{fig:our_results}
\end{figure*}

\begin{table}[!t]
    \centering
    \caption{\textbf{Comparison of different tokenization strategies.} The values for the naive method are shown on the left, while the values for our optimized method are shown on the right. $\star$ Inference time is tested on an RTX 4090 GPU. $\dagger$ indicates that the models were trained for only 160 epochs for this ablation study, to control for variables, so the results are not as good as full training. }
    \resizebox{\linewidth}{!}{
    \begin{tabular}{c|c|c|c}
    \hline
    \diagbox{Metrics}{Dataset} & Mixamo$^{\ast}$ & VRoid$^{\ast}$ & $\Dataset^{\ast}$ \\
    \hline
    Average Tokens & $369.53 \mid {\bf 214.89}$  & $621.76 \mid {\bf 522.88}$ & $495.46 \mid {\bf 237.94}$ \\
    \hline\hline
    Inference Time(s)$^{\star}$ & $3.57 \mid {\bf 2.16}$ & $5.39 \mid {\bf 4.53}$ & $4.29 \mid {\bf 1.99}$ \\
    \hline\hline
    J2J Distance$^{\dagger}$ & $0.1761 \mid {\bf 0.0838}$ & $0.1484 \mid {\bf 0.1374}$ & $0.1395 \mid {\bf 0.1266}$ \\
    J2B Distance$^{\dagger}$ & $0.1640 \mid {\bf 0.0779}$ & $0.1287 \mid {\bf 0.0891}$ & $0.1258 \mid {\bf 0.1017}$ \\
    B2B Distance$^{\dagger}$ & $0.1519 \mid {\bf 0.0715}$ & $0.1132 \mid {\bf 0.0766}$ & $0.1099 \mid {\bf 0.0966}$ \\ 
    \hline
    \end{tabular}
    }
    \label{tab:token_strategy}
    \vspace{-3mm}
\end{table}

\subsubsection{Skinning Weight Prediction and Mesh Deformation Robustness}\label{sec:skin_metric}

To evaluate the quality of our predicted skinning weights, we adopted a two-pronged approach: (1) \emph{direct comparison of skinning weights} and (2) \emph{evaluation of mesh deformation robustness under animation}. The former directly assesses the accuracy of the predicted weights, while the latter provides a more holistic measure of their ability to drive realistic animations.

For the \emph{direct comparison of skinning weights}, we computed the per-vertex L1 loss between the predicted and ground-truth skinning weights. We compared our method against RigNet \cite{xu2020rignet}, Neural Blend Shapes (NBS) \cite{li2021learning}, and TA-Rig \cite{ma2023tarig}, all of which also predict skinning weights. As shown in Table \ref{tab:skin_metric}, UniRig significantly outperforms these methods across all datasets, demonstrating the superior accuracy of our skin weight prediction.

As shown in Sections \ref{sec:bone_metric} and \ref{sec:skin_metric}, our method demonstrates substantial advantages in both skeleton rigging and skinning weight prediction, while also facilitating an efficient retargeting process. Consequently, the deformed meshes driven by our predictions exhibit good robustness across various animated poses. To quantify and validate this, 
we applied a set of 2,446 diverse animation sequences from the Mixamo dataset to the rigged models (VRoid and Mixamo). For each animation sequence, we sampled one frame and computed the L2 reconstruction loss between the ground-truth mesh and the mesh deformed using the predicted skeleton and skinning weights. This metric quantifies the ability of our method to produce realistic deformations across a wide range of motions.

Table \ref{tab:motion_metric} shows the reconstruction loss for UniRig and NBS. Our method achieves significantly lower reconstruction losses across all datasets, indicating its superior ability to generate robust and accurate mesh deformations. Notably, the results on ``VRoid with Spring*'' demonstrate the effectiveness of our method in handling dynamic simulations driven by spring bones.

Figure \ref{fig:compare_motion_commercial} provides a qualitative comparison of mesh deformation under motion against commercial tools (Meshy and Accurig) and NBS. The results demonstrate that our method produces more realistic deformations, particularly in areas with complex motion, such as the hair and hands. Figure \ref{fig:our_results} showcases the predicted skeletons, skinning weights, and resulting mesh deformations for various object types, further demonstrating the effectiveness of our approach.

\subsection{Ablation Study}

To validate the effectiveness of key components of our method, we conducted a series of ablation studies. Specifically, we investigated the impact of (1) our proposed tokenization strategy, (2) the use of indirect supervision via physical simulation, and (3) the training strategy based on skeletal equivalence.

\subsubsection{Tokenize Strategy}

In this comparative experiment, we assessed the performance of the naive tokenization method, as outlined in Section \ref{sec:ar-skeleton}, against our optimized approach. We evaluated both methods based on the following metrics: average token sequence length, inference time, and bone prediction accuracy (measured by J2J distances). For a fair comparison, both models were trained for 160 epochs. 
Table \ref{tab:token_strategy} shows the results of this comparison. Our optimized tokenization strategy significantly reduces the average token sequence length, leading to a decrease in inference time. Notably, it also improves bone prediction accuracy across all datasets, demonstrating the effectiveness of our approach in capturing skeletal structure. The inference time is tested on a single RTX 4090 GPU.

\subsubsection{Indirect Supervision based on Physical Simulation}

To evaluate the impact of indirect supervision using physical simulation (Section \ref{sec:physics}), we compared the performance of our model with and without this component during training. We focused on the VRoid dataset for this experiment, as it contains spring bones that are directly affected by the physical simulation. Table \ref{tab:physical_strategy} shows that training with indirect supervision leads to a significant improvement in both deformation error (L2 loss) and skinning weight error (L1 loss). This demonstrates that incorporating physical simulation into the training process helps the model learn more realistic skinning weights and bone attributes. 

\begin{table}[!t]
    \centering
    \caption{Ablation study on the use of indirect supervision via physical simulation. Deformation error is tested using the L2 loss under the same motion, while skinning error is evaluated using the L1 loss of per-vertex skinning weights.}
    \vspace{-2mm}
    \resizebox{\linewidth}{!}{
        \begin{tabular}{c|c|c}
        \hline
        \diagbox[width=3.6cm]{Method}{Metrics} & Deformation Error & Skin Error\\
        \hline
        \Method & $\bf 7.74 \times 10^{-4}$ & $\bf 5.42 \times 10^{-3}$\\
        w/o Physical Simulation & $8.59 \times 10^{-4}$ & $5.78 \times 10^{-3}$\\
        \hline
        \end{tabular}
    }
    \vspace{-2mm}
    \label{tab:physical_strategy}
\end{table}

\subsubsection{Training Strategy Based on Skeletal Equivalence}

To validate the effectiveness of our training strategy based on skeletal equivalence (Section \ref{sec:skin_pred}), we compared the performance of our model with and without this strategy. Specifically, we evaluated the impact of two key components: (1) randomly freezing bones during training and (2) normalizing the loss by the number of influenced vertices for each bone.
Table \ref{tab:skin_strategy} shows the results of this comparison. Using the full skeletal equivalence strategy (\emph{UniRig}) yields the best performance in terms of reconstruction loss. Disabling either component (``w/o skeleton frozen'' or ``w/o bone loss normalization'') leads to a degradation in performance, highlighting the importance of both aspects of our training strategy in achieving optimal results.

\begin{table}[!t]
    \centering
    \caption{Ablation study on the training strategy based on skeletal equivalence. $\star$ indicates that the evaluation dataset is under the data augmentation of random rotation, scale, and applying random motion.}
    \resizebox{\linewidth}{!}{
        \begin{tabular}{c|c|c|c}
        \hline
        \diagbox[width=3.7cm]{Metrics}{Dataset} & Mixamo$^{\ast}$ & VRoid$^{\ast}$ & $\Dataset^{\ast}$ \\
        \hline
        \Method & $\bf 4.42 \times 10^{-4}$ & $1.28 \times 10^{-3}$ & $\bf 3.72 \times 10^{-3}$\\
        w/o skeleton frozen  & $4.92 \times 10^{-4}$  & $\bf 1.25 \times 10^{-3}$  & $3.84 \times 10^{-3}$ \\
        w/o bone loss normalization & $4.63 \times 10^{-4}$  & $1.33 \times 10^{-3}$ & $3.92 \times 10^{-3}$ \\
        \hline
        \end{tabular}
    }
    \label{tab:skin_strategy}
    \vspace{-4mm}
\end{table}

\section{Applications\label{sec:Application}}

\subsection{Human-Assisted Auto-rigging\label{sec:human_edit}}

\begin{figure}[!t]
    \centering
    \includegraphics[width=0.95\linewidth]{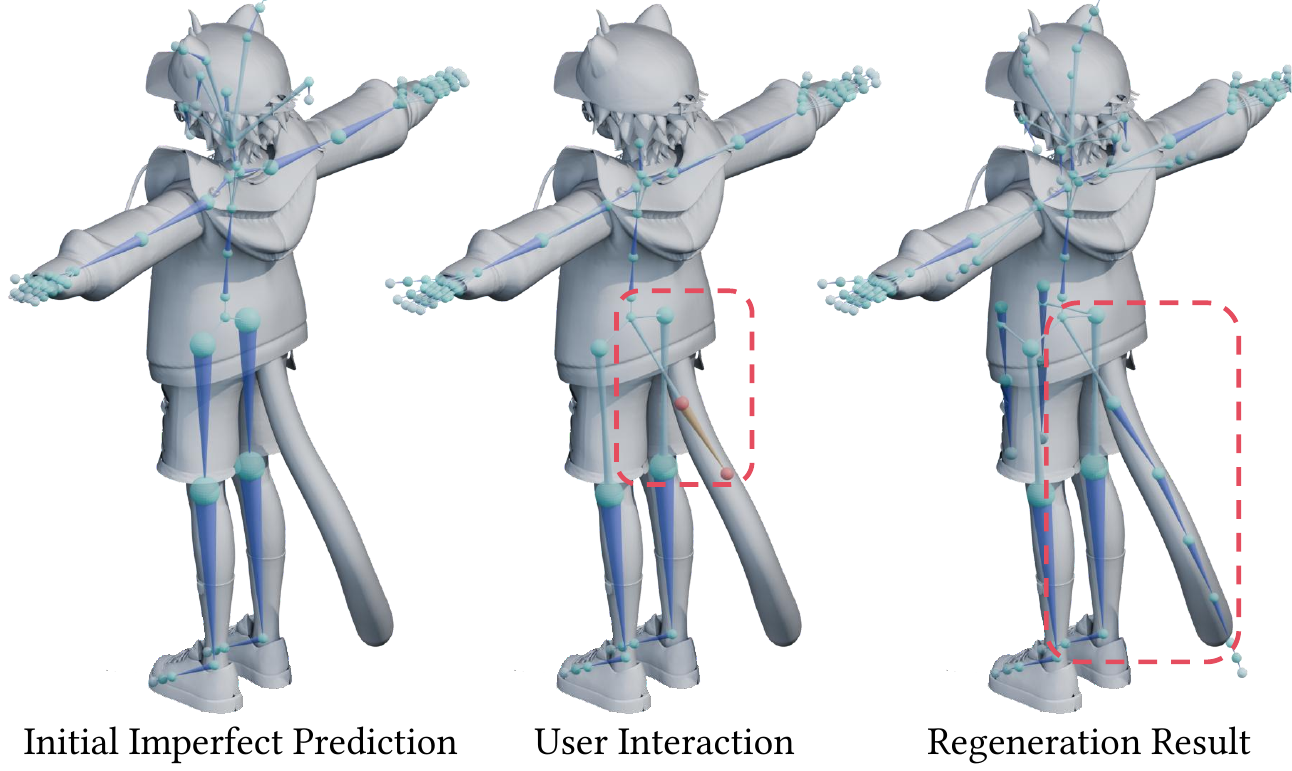}
    \caption{\textbf{Human-assisted skeleton editing and regeneration with \Method.} In this example, the initial prediction lacks a tail and has unsatisfactory spring bones. The user removes the spring bones, keeps the Mixamo template skeleton, and adds a prompt for a tail bone. \Method then regenerates the skeleton based on these modifications, resulting in a more accurate and desirable rig.}
    \label{fig:edit_method}
\end{figure}

Compared to prior automatic rigging techniques, a key advantage of our approach lies in its ability to facilitate human-machine interaction. This is achieved through the ability to edit the predicted skeleton tree and trigger subsequent regeneration of the affected parts. As shown in Figure \ref{fig:edit_method}, users can perform operations such as adding new bone branches or removing existing ones (e.g., removing spring bones to achieve a more rigid structure). This allows for efficient correction of any inaccuracies in the automatic prediction and customization of the rig to specific needs. For instance, a user might add a new branch to represent a tail that was not automatically detected, or they might remove automatically generated spring bones that are not desired for a particular animation. The edited skeleton tree can then be fed back into the \Method pipeline, generating an updated rig that incorporates the user's modifications. This iterative process empowers users to quickly and easily refine the automatically generated rigs, combining the speed of automation with the precision of manual control.

\subsection{Character Animation}
\begin{figure}[!t]
    \centering
    \includegraphics[width=0.7\linewidth]{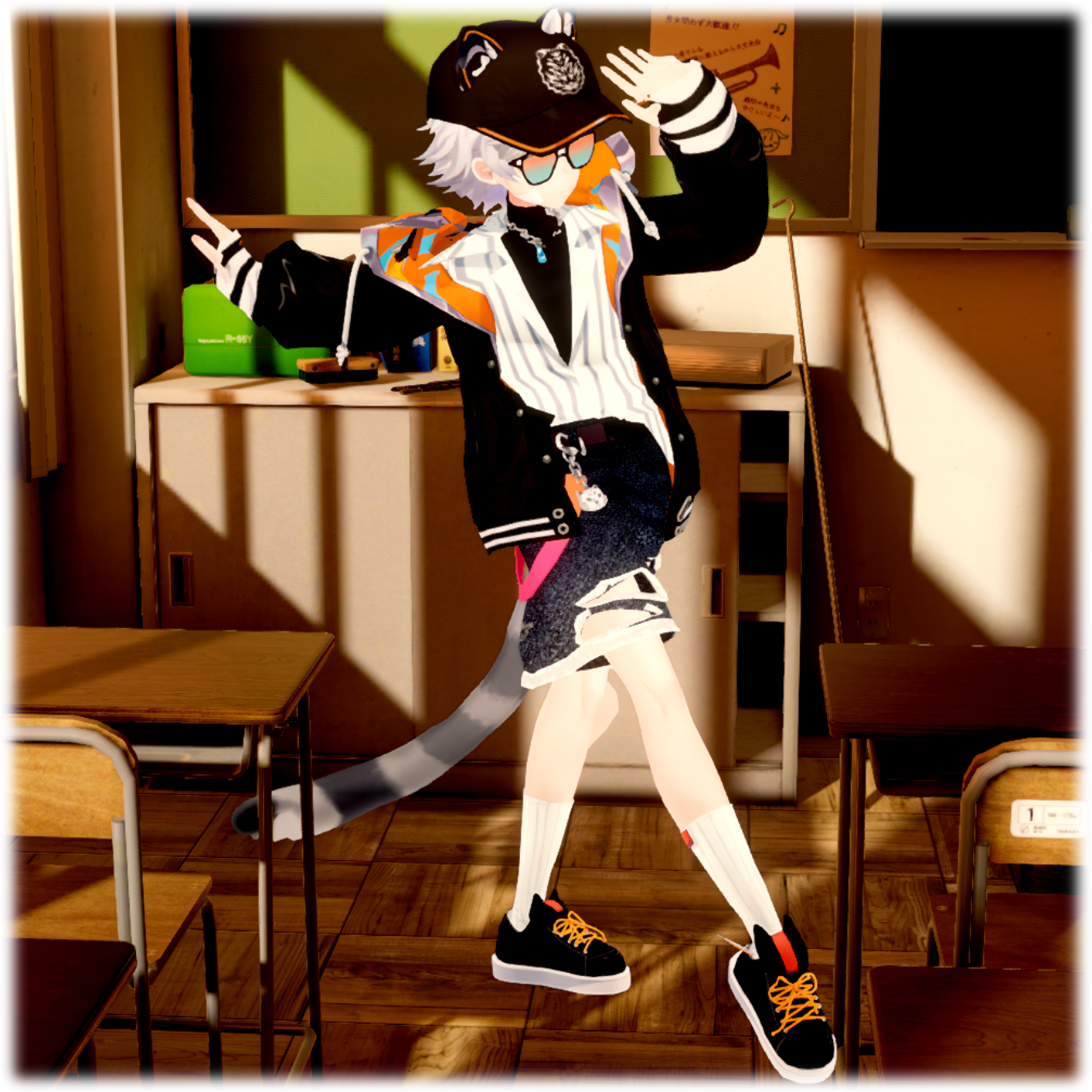}
    \caption{\textbf{VTuber live streaming with a UniRig-generated model.} The character, rigged using our method, exhibits smooth and realistic spring bone motion during live streaming in Warudo \cite{tang2024warudo}.}
    \label{fig:warudo}
    \vspace{-5mm}
\end{figure}

\emph{UniRig}'s ability to predict spring bone parameters, trained on the VRoid and Rig-XL dataset, makes it particularly well-suited for creating animated characters. Our method can generate VRM-compatible models from simple mesh inputs, enabling users to easily export their creations to various animation platforms. This streamlines the process of creating and animating virtual characters. For example, users can leverage tools like Warudo \cite{tang2024warudo} to bring their rigged characters to life in a virtual environment, as demonstrated in Figure \ref{fig:warudo}. This capability is especially valuable for applications like VTubing, where realistic and expressive character motion is highly desirable. The smooth and natural movements generated by our spring bone simulation contribute to a more engaging and immersive VTubing experience.

\section{Conclusions\label{sec:limitation}}

This paper presents \Method, a unified learning-based framework for automatic rigging of 3D models. Our model, combined with a novel tokenization strategy and a two-stage training process, achieves state-of-the-art results in skeleton prediction and skinning weight prediction. The large-scale and diverse \Dataset dataset, along with the curated VRoid dataset, enables training a generalizable model that can handle a wide variety of object categories and skeletal structures. 

\noindent \textbf{Limitations and Discussions.} Despite its strengths, \Method has certain limitations. Like other learning-based approaches, the performance of our method is inherently tied to the quality and diversity of the training data. While \Dataset is a large and diverse dataset, it may not fully encompass the vast range of possible skeletal structures and object categories. Consequently, \Method might perform suboptimally when presented with objects that significantly deviate from those in the training data. For instance, it might struggle with highly unusual skeletal structures, such as those found in abstract or highly stylized characters. As mentioned in Section \ref{sec:human_edit}, user edits can be used as a valuable source of data for further refining the model. By incorporating user feedback and expanding the training dataset, we can continuously improve the robustness and generalizability of \Method. There are several avenues for future work. One direction is to explore the use of different modalities, such as images or videos, as input to the rigging process. Furthermore, incorporating more sophisticated physical simulation techniques could enhance the realism of the generated animations.

In conclusion, \Method represents a step towards fully automated and generalizable rigging. Its ability to handle diverse object categories, coupled with its support for human-in-the-loop editing and realistic animation, makes it a powerful tool for both researchers and practitioners in the field of 3D computer graphics.

\bibliographystyle{ACM-Reference-Format}
\bibliography{reference}

\clearpage
\appendix

\section{Appendix}

\begin{table*}[!t]
    \centering
    \caption{Joint to bone (J2B) and Bone to bone (B2B) Chamfer distance. Left is CD-J2B, and right is CD-B2B. $^{\ast}$ means the evaluation dataset is under the data augmentation of random rotation, scale and applying random motion. $^\dagger$ means we cannot finetune the model because RigNet do not provide data preprocess tools and TA-Rig do not provide training scripts.}
    \resizebox{\linewidth}{!}{
    \begin{tabular}{l|c|c|c|c|c}
    \hline
      \diagbox[width=4.5cm]{Method}{Dataset}   & Mixamo & VRoid & Mixamo$^{\ast}$ & VRoid$^{\ast}$ &  $\Dataset^{\ast}$ \\
         \hline
      Ours  & $\bf{0.0077 \mid 0.0044}$ & $\bf{0.0076 \mid 0.0043}$ & $\bf{0.0075 \mid 0.0040}$ & $\bf{0.0085 \mid 0.0046}$ & $\bf{0.0456 \mid 0.0276}$ \\
      RigNet$^\dagger$ \cite{xu2020rignet} & $0.0470 \mid 0.0398$  & $0.1992 \mid 0.1793$  & $0.1719 \mid 0.1534$ & $0.2082 \mid 0.1833$ & $0.1847 \mid 0.1519$ \\
      Neural Blend-Shape\cite{li2021learning} & $0.0277 \mid 0.0181$  & $0.0158 \mid 0.0108$ & $0.0349 \mid 0.0232$ & $0.0168 \mid 0.0113$ &  N/A \\
      TA-Rig$^\dagger$ \cite{ma2023tarig}  & $0.0937 \mid 0.0775$  & $0.0832 \mid 0.0682$ & $0.1027 \mid 0.0860$ & $0.0884 \mid 0.0726$ & $0.1892 \mid 0.1465$ \\
      \hline
    \end{tabular}
    }
    \label{tab:bone_metric_jb2b}
\end{table*}

\begin{table*}[!t]
    \centering
    \caption{Quantitative comparison of skeleton prediction on
Model Resources-RigNet\cite{Models-Resource, xu2020rignet}.}
    \resizebox{0.7\linewidth}{!}{
    \begin{tabular}{l|c|c|c|c|c}
    \hline
      \diagbox[width=4.5cm]{Method}{Metrics}   & CD-J2J & CD-J2B & CD-B2B & Skin L1 & Motion L2 \\
         \hline
      Ours  & $\bf{0.0332}$ & $0.0266$ & $\bf{0.0194}$ & $\bf{0.0455}$ & $\bf{0.0019}$ \\
      RigNet$^\dagger$\cite{xu2020rignet} & $0.039$  & $\bf{0.024}$  & $0.022$ & $0.39$ & N/A \\
      Anything World & $0.0540$  & $0.0528$ & $0.0338$ & N/A & N/A \\
      \hline
    \end{tabular}
    }
    \label{tab:your_table_name}
\end{table*}

\subsection{Datasets}

\subsubsection{\Dataset Data Process}

\paragraph{Fix the problem of lacking a reasonable topological relationship.\label{appendix:topo_mistake}}{
When processing Objaverse, we found that many animators do not rig a reasonable topology, because sometimes they directly use keyframe animation to adjust the bones individually to create the animation. This situation can be filtered by a simple rule: if the out-degree of the root node is greater than $4$, and the subtree size of the root node's \textbf{heavy child} exceeds half the size of the skeleton Tree, the vast majority of such data can be filtered out. To address this issue, we cut off all outgoing edges of the root node, treat the \textbf{heavy child} as the new root, and then connect the remaining forest using a minimum spanning tree(MST) based on Euclidean distance.}

\subsection{More filter rules about the \Dataset}

\subsubsection{Capture outlier through reconstruction loss}

In the blend skinning weight training in Section \ref{sec:skin_pred}, we found that although many data points were filtered, there were still a few outliers in the reconstruction loss. This is actually because there were still some non-compliant data that were not cleared during the Objaverse data preprocessing. Therefore, we used the current average reconstruction loss multiplied by 10 as a threshold and filtered out the incorrectly preprocessed data during multiple epochs of training, removing it from the dataset.
In addition, we removed samples where the skinning weights of some points were completely lost, because softmax is applied on each point, which makes it impossible to fit situations where all weights of the point are zero.

\subsection{Methods}

\subsubsection{Physical Simulation on VRM}

When deforming the VRM body, it first calculates the basic motion of the body using the forward kinematics method (i.e., the standard Mixamo template). Then, for each spring bone, the Verlet integration is applied sequentially from top to bottom along the chain to compute the position of each spring bone, resulting in a coherent animation effect. Whole process is shown in Algorithm \ref{code:physics}.

\begin{algorithm}[!t]
\caption{Verlet Integration for Bone Position Update}
\KwIn{$T_{\text{current}}$: Bone tail of current frame, $T_{\text{prev}}$: Bone tail of previous frame, $L_{\text{bone}}$: Bone length, $\eta_d$: Drag coefficient, $\eta_s$: Stiffness coefficient, $\eta_g$: Gravity coefficient, $g$: Gravity direction, $\Delta t$: Time step.}
\KwOut{$T_{\text{next}}$: Updated bone tail position of the next frame.}

\label{code:physics}

\SetKwFunction{FUpdate}{UpdatePosition}
\SetKwProg{Fn}{Function}{:}{end}

\Fn{\FUpdate{$T_{\text{current}}, T_{\text{prev}}, L_{\text{bone}}, \eta_d, \eta_s, \eta_g, g, \Delta t$}}{
    ${\bf I} \gets (T_{\text{current}} - T_{\text{prev}}) \cdot (1 - \eta_d)$; \tcp{Calculate interia}

    ${\bf S} \gets \eta_s R_{\text{head}}^{-1} R_{\text{tail}}$; \tcp{Calculate stiffness, $R$ is the rotation matrix under world coordinate system}

    ${\bf G} \gets \eta_g \cdot \bf g$; \tcp{Calculate gravity}

    $\Delta x \gets {\bf (I + S + G)} \cdot \Delta t$; \tcp{Calculate displacement of the bone tail under three forces}

    $T_{\text{next}} \gets H_{\text{next}} + L_{\text{bone}} \displaystyle \frac{\Delta x}{|\Delta x|}$ \tcp{Update next tail position under length normalization}
    \KwRet{$T_{\text{next}}$}\;
}
\end{algorithm}

\begin{figure}[!b]
    \centering
    \includegraphics[width=\linewidth]{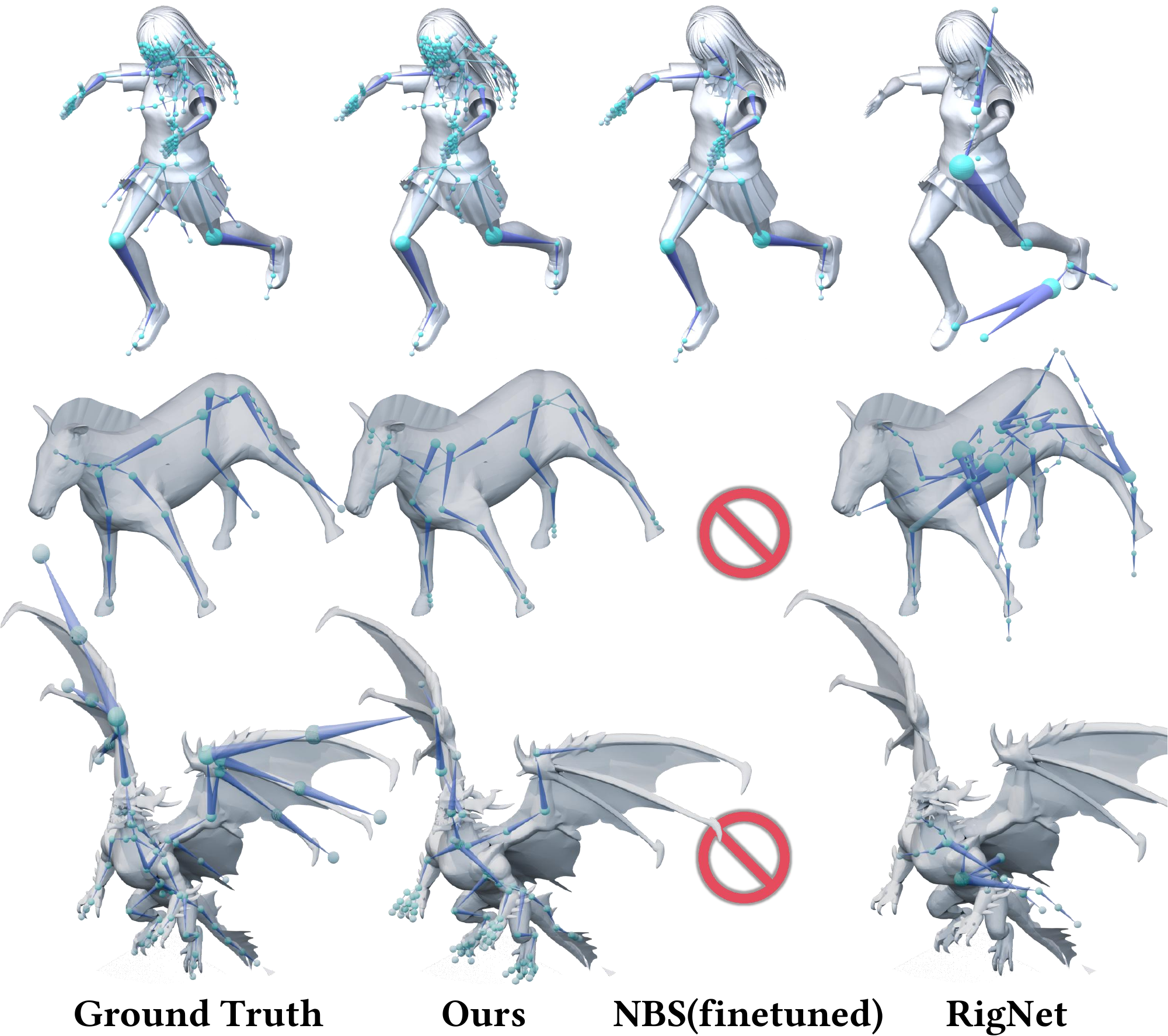}
    \caption{We compare auto-rigging skeleton with NBS(finetuned) and RigNet on different kinds of 3D models.}
    \label{fig:compare_skeleton_academic}
\end{figure}

\begin{figure}[!b]
    \centering
    \includegraphics[width=\linewidth]{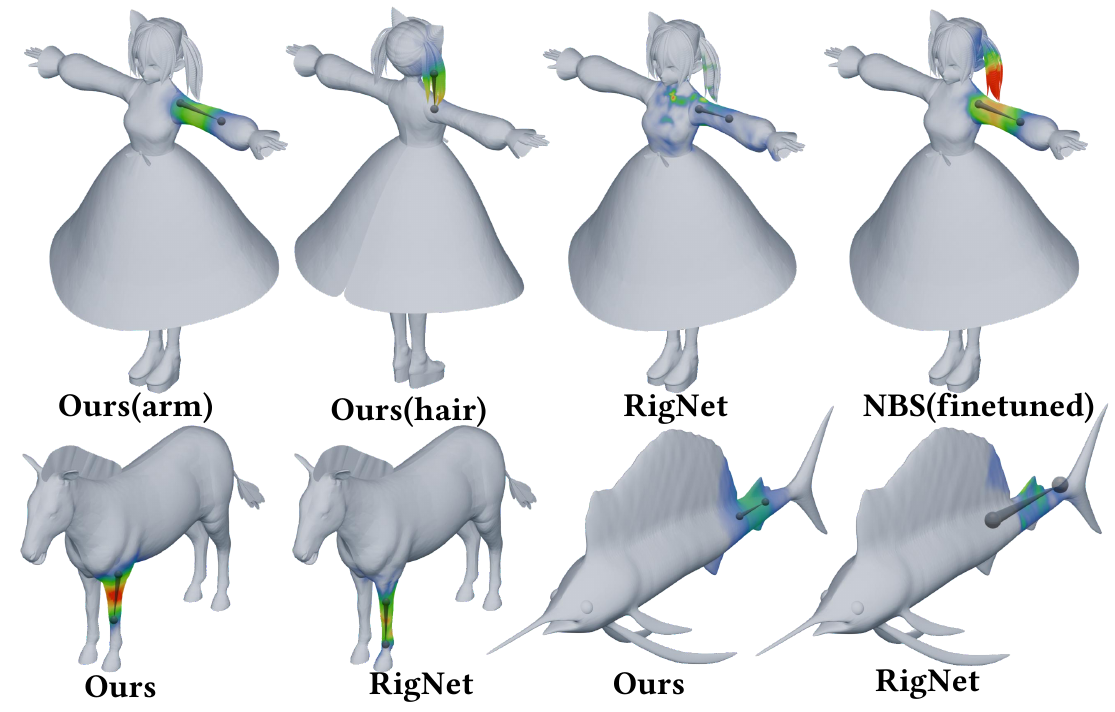}
    \caption{We compare blend skinning weight with NBS(finetuned) and RigNet on different kinds of 3D models.}
    \label{fig:compare_skin_academic}
\end{figure}

\begin{figure*}[!b]
    \centering
    \includegraphics[width=0.9\linewidth]{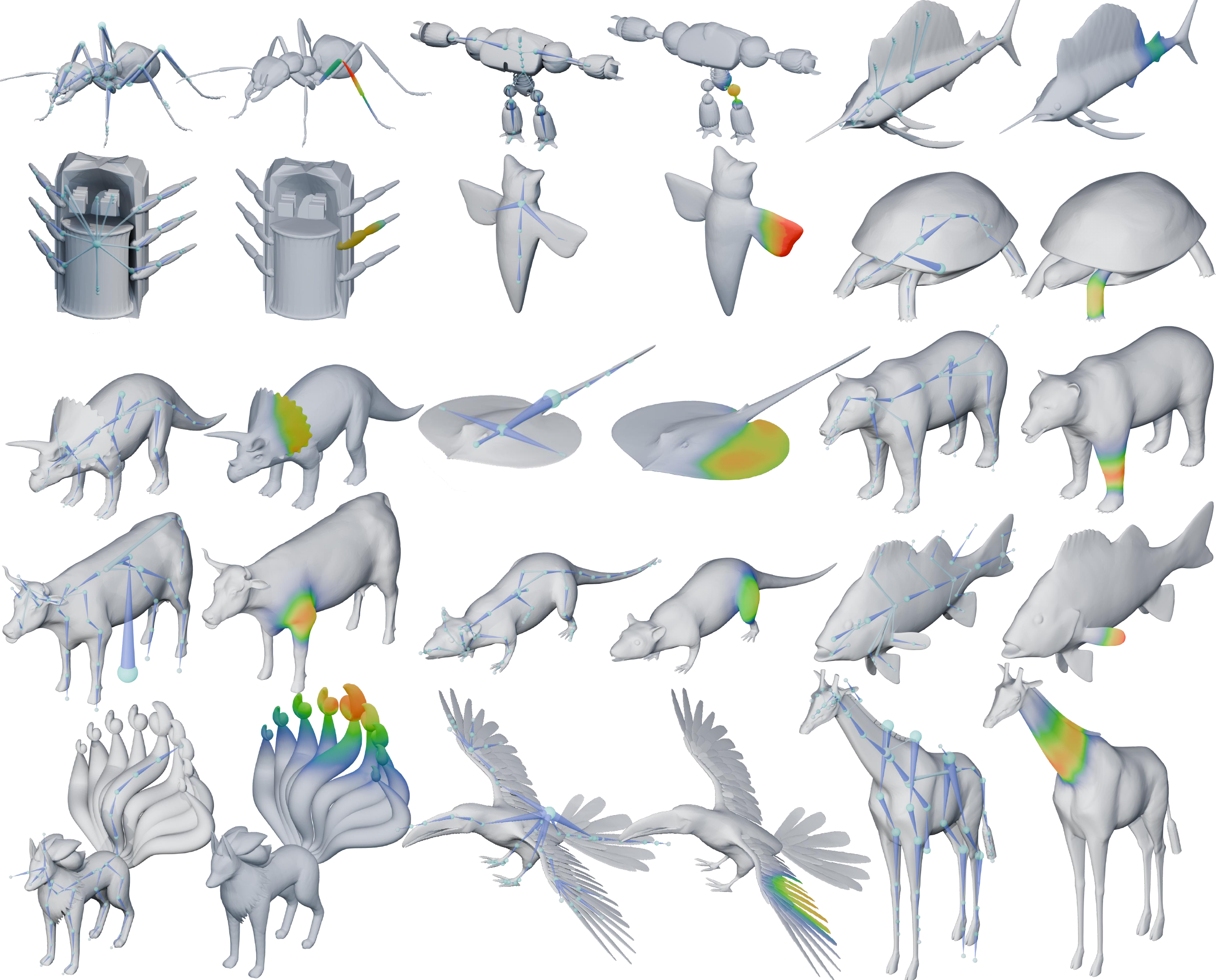}
    \caption{We present more examples of \Method here, demonstrating highly detailed and accurate skeleton rigging and weight generation.}
    \label{fig:more_result}
\end{figure*}

We show more visualization results for detailed comparison. 
In Figure \ref{fig:compare_skeleton_academic}, we compare \Method with NBS and RigNet on different types of examples for automatic rigging, which can be observed that it can predict highly accurate and detailed results even for non-standard poses and various complex meshes. 
Figure \ref{fig:compare_skin_academic} demonstrates the precision of \Method in predicting skinning weights such as hair better than previous work.
Finally, Figure \ref{fig:more_result} showcases the high-precision skeleton rigging and excellent weight generated achieved by \Method on more complex examples, such as ants.

\subsection{More Results}

\end{document}